\documentclass[amsmath,amsfonts,amssymb,twocolumn,aps,prl,preprintnumbers,superscriptaddress]{revtex4-2}

\usepackage[utf8]{inputenc}
\usepackage[T1]{fontenc}
\usepackage{lingmacros} 
\usepackage{tree-dvips} 
\usepackage{graphicx}
\usepackage{tabularx}
\usepackage{multirow}
\usepackage{dcolumn}
\usepackage{booktabs}
\usepackage{bm}
\usepackage{color}
\usepackage{textcomp} 

 \usepackage[normalem]{ulem} 

\definecolor{red}{rgb}{0.8,0,0.1}

\definecolor{darkblue}{rgb}{0,0,.6}

\definecolor{lightgrey}{rgb}{0.7,0.7,0.7}

\definecolor{grey}{rgb}{0.4,0.4,0.4}

\newcommand*{\Tx}{\ensuremath{T_{\mathrm{exc}}}~}

\begin{document}

\title{Spectral asymmetry of phonon sideband luminescence in monolayer and bilayer WSe$_2$}

\def\LMU{Fakult\"at f\"ur Physik, Munich Quantum Center,
  and Center for NanoScience (CeNS),
  Ludwig-Maximilians-Universit\"at M\"unchen,
  Geschwister-Scholl-Platz 1, 80539 M\"unchen, Germany}
\def\Regensburg{Departement of Physics, University of Regensburg,  93053 Regensburg, Germany}
\def\Dresden{Institute for Applied Physics, Dresden University of Technology, Dresden, 01187, Germany}
\def\TsukubaFunctional{ Research Center for Functional Materials, National Institute for Materials Science, Tsukuba 305-0044, Japan}
\def\TsukubaNano{International Center for Materials Nanoarchitectonics, National Institute for Materials Science, Tsukuba 305-0044, Japan}

\author{Victor Funk}
\affiliation{\LMU}
\author{Koloman Wagner}
\affiliation{\Regensburg}
\author{Edith Wietek}
\affiliation{\Regensburg}
\author{Jonas D. Ziegler}
\affiliation{\Regensburg}
\author{Jonathan F\"orste}
\affiliation{\LMU}
\author{Jessica Lindlau}
\affiliation{\LMU}
\author{Michael F\"org}
\affiliation{\LMU}
\author{Kenji Watanabe}
\affiliation{\TsukubaFunctional}
\author{Takashi Taniguchi}
\affiliation{\TsukubaNano}
\author{Alexey Chernikov}
\affiliation{\Regensburg}
\affiliation{\Dresden}
\author{Alexander H\"ogele}
\affiliation{\LMU}
\affiliation{Munich Center for Quantum Science and Technology (MCQST),
  Schellingtra\ss{}e 4, 80799 M\"unchen, Germany}

\date{\today}

\begin{abstract}
We report an experimental study of temperature-dependent spectral lineshapes of phonon sideband emission stemming from dark excitons in monolayer and bilayer WSe$_2$. Using photoluminescence spectroscopy in the range from $4$ to $100$~K, we observe a pronounced asymmetry in the phonon-assisted luminescence from momentum-indirect exciton reservoirs. We demonstrate that the corresponding spectral profiles are distinct from those of bright excitons with direct radiative decay pathways. The lineshape asymmetry reflects thermal distribution of exciton states with finite center-of-mass momenta, characteristic for phonon sideband emission. The extracted temperature of the exciton reservoirs is found to generally follow that of the crystal lattice, with deviations reflecting overheated populations. The latter are most pronounced in the bilayer case and at lowest temperatures. Our results add to the understanding of phonon-assisted recombination of momentum-dark excitons and, more generally, establish means to access the thermal distribution of finite-momentum excitons in atomically thin semiconductors with indirect bandgaps.
\end{abstract}

\newpage

\maketitle
The nature of radiative electron-hole transitions has a rich history in semiconductor physics. Already from the early studies of germanium and silicon it became apparent that the requirements of simultaneous energy and momentum conservation fundamentally determine the optical response of semiconductors in both absorption \cite{MacfarlaneGe,MacfarlaneSi} and emission \cite{Haynes1952,Newman1953,HaynesGe,HaynesSi}. Spectroscopy thus provides convenient and powerful means for discriminating between momentum-direct and momentum-indirect transitions, as well as for identifying interface- and phonon-assisted recombination \cite{Bonfanti2008, Giorgioni2012}. Related methods have been successfully applied to two-dimensional van der Waals semiconductors to identify the prominent indirect-to-direct crossover from the bulk to the monolayer (ML) limit of transition metal dichalcogenides (TMDs) \cite{Mak2010,Splendiani2010}. 

TMDs offer a particularly viable platform to study the physics of momentum-direct and indirect electron-hole recombination processes in one material system. In contrast to conventional semiconductors, their properties are strongly affected by sizable Coulomb interactions, leading to the formation of tightly-bound exciton quasiparticles \cite{Wang2018} with intriguing spin-valley physics resulting from the underlying band structure \cite{Xu2014}. Following recent advances in sample preparation and doping control \cite{Mak2013,Ross2013,Cadiz2017,Ajayi2017,Wierzbowski2017}, TMDs exhibit a rich spectrum of exciton photoluminescence (PL) associated with electron-hole pairs from different high-symmetry points of the first Brillouin zone.

The PL of charge-neutral few-layer TMDs can be generally grouped into features of excitons with zero center-of-mass momentum \cite{Mak2010,Splendiani2010,Zhou2017,Zhang2017} and phonon sidebands of dark excitons with finite momenta \cite{Lindlau2017ML,Lindlau2018,Brem2020,He2020}. The former decay radiatively with in- or out-of-plane polarization resulting in characteristically symmetric spectral profiles in PL. Contrary to that, the latter originate from higher-order processes that require the assistance of phonons to conserve momentum upon photon emission. In the specific case of WSe$_2$, phonon sidebands enrich the ML emission spectra by adding peaks energetically below the bright exciton X \cite{Arora2015,Courtade2017,Zhou2017,Zhang2017,Lindlau2017ML,Koperski2017,LiuPRR2019,He2020,Liu2020,Forste2020}. The respective lower-energy PL features stem from $KK^\prime$ and $KQ$ excitons and correspond to phonon-assisted recombinations between valence band vacancies in the $K$ valley and conduction band electrons in the $K^\prime$ or $Q$ valley, respectively. In WSe$_2$ bilayers (BLs), the energy of zero-momentum $KK$ excitons remains energetically close to X of the ML, whereas the energy of $KQ$ excitons is substantially reduced due to interlayer hybridization effects \cite{Mak2010,Splendiani2010}. As a result, the $KQ$ exciton reservoirs host the majority of the photoexcited exciton population at low temperatures and thus dominate the cryogenic PL spectra of BL WSe$_2$ \cite{Lindlau2018,Aslan2020,Forste2020}.
\begin{figure}[t]
\centering
\includegraphics[scale=1.0]{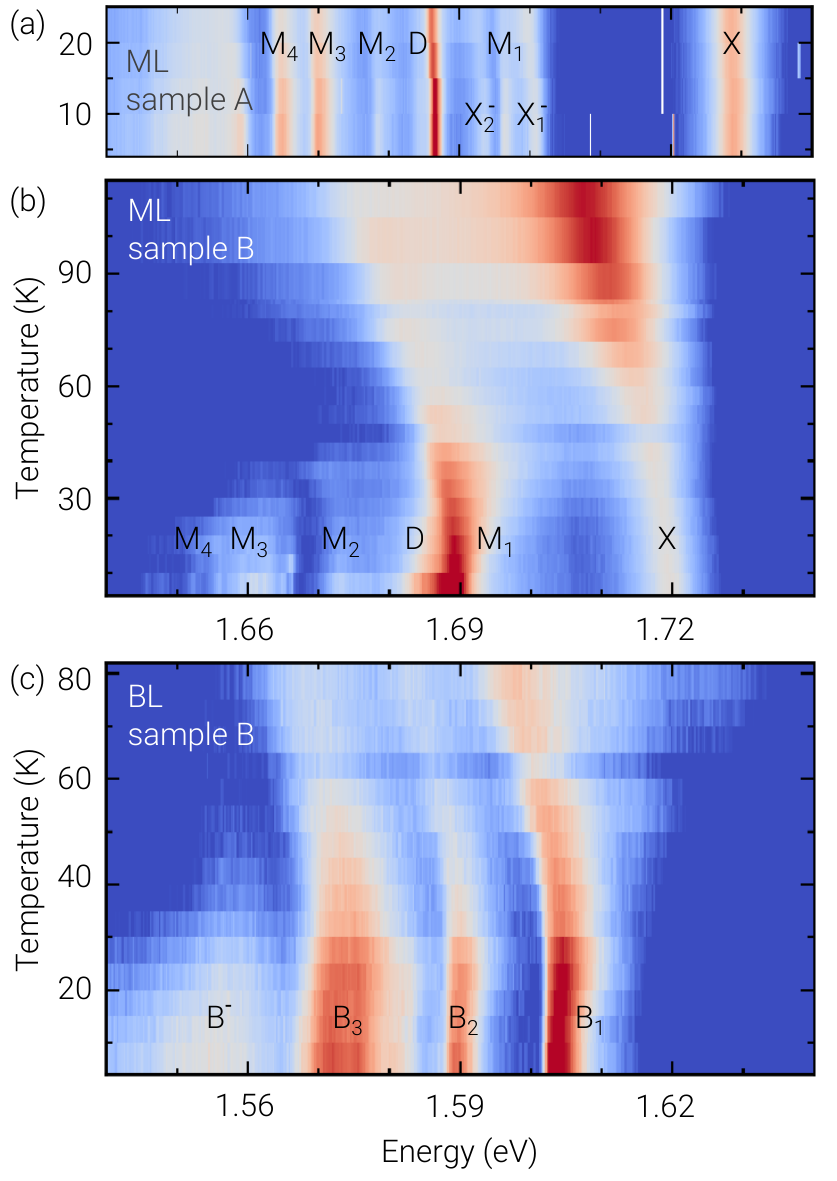}
\caption{(a), (b) Logarithmic false-color plots of the PL from samples A and B of ML WSe$_2$ encapsulated in hBN, where the former exhibits features of residual doping and the latter is tuned to charge neutrality. (c) PL of BL WSe$_2$ from sample B tuned close to charge neutrality with a marginal contribution of the negative bilayer trion B$^-$. Neutral spin-bright and spin-dark $KK$ exciton peaks are denoted as X and D, and the trion doublet as X$_{1}^-$ and X$_{2}^-$. The remaining ML and BL peaks are labeled by an increasing subscript number $n$ as M$_n$ and B$_n$, respectively.}
\label{fig1}
\end{figure}

\begin{figure*}[t]
\centering
\includegraphics[scale=1.0]{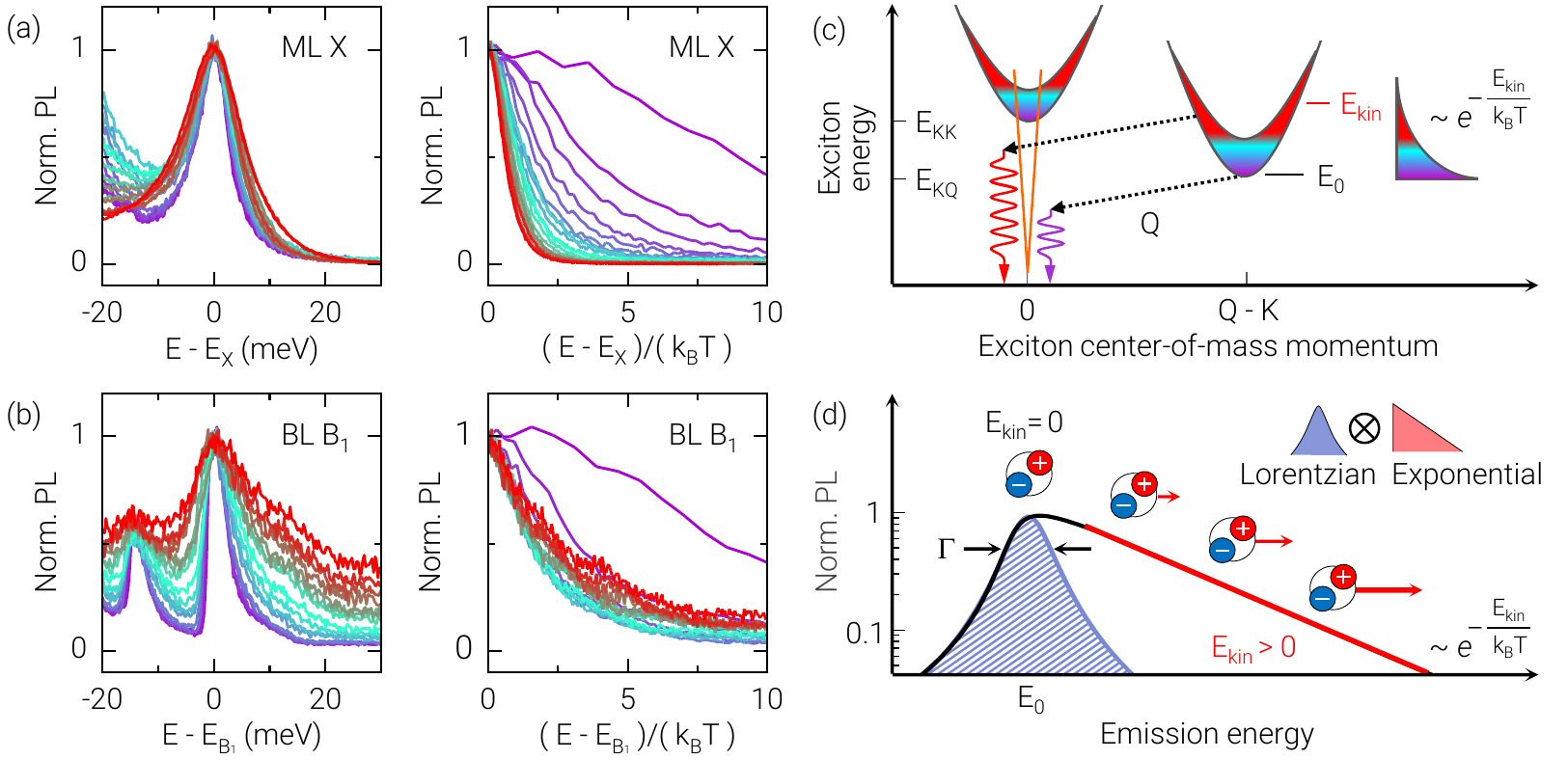}
\caption{(a) Left panel: PL spectra of ML in sample B at temperatures from $4$ to $120$~K (colored from purple to red), normalized to unity and shifted horizontally to the energy $E_X$ of the bright exciton peak X at $4$~K. Right panel: High-energy shoulder of the spectra with energy axis normalized by $k_B T$. (b) Same for BL in the temperature range from $4$ to $80$~K with the energy of the peak B$_1$ as reference. (c) Schematic representation of direct and indirect decay processes for excitons. The center-of-mass momenta of $KK$ excitons, such as X or D, with zero-phonon radiative decay are restricted to 'cold' excitons (blue shaded population inside the orange light cone). In contrast, phonon sideband luminescence of momentum-dark excitons such as $KQ$ with energy minimum $E_0=E_{KQ}$ and exciton center-of-mass momentum $Q-K$ is assisted by phonons with momentum $Q$ and thus also includes emission from 'hot' excitons (thermally activated red shaded population). The thermal Boltzmann occupancy of exciton states with finite kinetic energy $E_{\mathrm{kin}}>E_0$ above their respective dispersion minimum is represented by the area of colored shades. (d) Illustration of the phonon sideband asymmetry: The homogeneously broadened Lorentzian spectrum with linewidth $\Gamma$ is convoluted on the high-energy side with the contribution from excitons with finite kinetic energy $E_{\mathrm{kin}}>0$ according to their thermal occupancy in Boltzmann approximation}.
\label{fig2}
\end{figure*}

To date, momentum-dark excitons and their phonon sidebands in WSe$_2$ have been considered in the context of PL yield and dynamics \cite{Moody2015,Selig2016,Malic2018,Selig2018,Berghaeuser2018,Peng2019,Rosati2020} rather than in terms of spectral lineshape analysis. In particular, considerations regarding the spectral form of the emission that should reflect the characteristics of excitons with finite momenta \cite{Klingshirn2007,Brem2020} have been largely neglected so far despite the fact that phonon-assisted emission would allow for direct optical access to the exciton distribution beyond the restrictive limit of the radiative cone that otherwise applies for zero-phonon transitions. In the present work, we address these topics and report temperature-dependent studies of profound differences in the lineshapes of PL peaks characteristic for direct and phonon-assisted transitions in ML and BL WSe$_2$. Our analysis demonstrates that asymmetric phonon sidebands are indeed captured by accounting for the thermal exciton distribution. In addition, we show that while the resulting effective exciton temperature generally follows that of the lattice, there are also notable deviations due to overheated exciton populations particularly pronounced in the BL case. 

The samples studied in this work are based on exfoliated van der Waals heterostructures consisting of ML (samples A and B) and BL (also on sample B) WSe$_2$ (HQ Graphene) encapsulated between hexagonal boron nitride (hBN) layers and deposited on a SiO$_2$/Si substrate by polymer-assisted all-dry viscoelastic stamping \cite{Castellanos-Gomez2014,Pizzocchero2016}. Subsequent annealing was carried out to enhance interlayer contact and reduce trapped residues \cite{Taniguchi2018}. Sample A was mounted in a continuous-flow cryostat in a confocal microscope and excited by a continuous-wave (cw) laser at $2.33$~eV with a power of $500$~nW into an optical spot of $1~\mu$m diameter. The PL was dispersed with a spectrometer and detected by a charge-coupled-device camera.

In the ML PL of sample A, shown in Fig.~\ref{fig1}(a) in the low-temperature range, the peak at $1.728$~eV stems from the neutral exciton X, followed by the weaker trion doublet, X$_{1}^-$ and X$_{2}^-$, with $6$~meV exchange splitting \cite{Courtade2017}. The doublet is present due to residual n-doping of the sample. Between the doublet peaks, the M$_1$ peak at $30$~meV red-shift from X has a rich history of controversial assignments including defect-activated recombination of $KK'$ momentum-dark intervalley excitons \cite{He2020,Liu2020}, phonon-assisted emission PL from virtual trions \cite{VanTuan2019}, and phonon sideband of momentum-dark $KQ$ excitons \cite{Lindlau2017ML,Forste2020}. A series of red-shifted peaks below the spin-dark $KK$ exciton (D) \cite{Zhang2015,Wang2017,Robert2017,Zhou2017}, labeled  by an increasing subscript number with decreasing peak energy as M$_2$ through M$_4$, represent phonon sideband luminescence from long-lived dark excitons. In charge-neutral samples, recent work has ascribed the peaks M$_2$ and M$_3$ at $51$ and $59$~meV red-shifts from X to phonon sidebands of momentum-dark $KK'$ excitons \cite{He2020,Liu2020}. However, it has been also shown that excitons from $KQ$ should exhibit phonon-assisted emission around $M_2$ \cite{Brem2020,Rosati2020}. Finally, the M$_4$ peak with $65$~meV red-shift from X corresponds to the chiral-phonon replica of long-lived spin-dark D excitons \cite{Li2019,LiuPRR2019,He2020,Liu2020}.

Sample B with data in Fig.~\ref{fig1}(b) and (c) was studied in a closed-cycle magneto-cryostat with a base temperature of $3.2$~K. Charge-doping in ML and BL was controlled via the field effect by grounding a gold electrode in contact with the TMD crystal and applying a gate voltage with respect to the highly doped silicon substrate \cite{Forste2020}. The sample was excited with a cw diode laser at $1.95$~eV at a low pump density in the linear regime. The temperature-dependent PL of a ML region on sample B tuned to charge neutrality is shown in Fig.~\ref{fig1}(b). At $3.2$~K, the PL characteristics are equivalent to those of sample A in Fig.~\ref{fig1}(a) with an overall sample-specific red-shift of $8$~meV, fully suppressed trion features and a larger inhomogeneous line broadening. For all peaks, a characteristic temperature-dependent red-shift was observed with increasing temperature \cite{Arora2015}. In addition, the PL intensity of the bright exciton X increased with temperature at the expense of all lower-lying peaks. This is consistent with a temperature-dependent population redistribution from lower dark to energetically higher bright exciton states, in agreement with previous experimental \cite{Zhang2015,Robert2017} and more recent theoretical studies \cite{Brem2020}.

Complementary to the ML results presented in Fig.~\ref{fig1}(a) and (b), temperature-dependent PL from encapsulated BL WSe$_2$ tuned close to charge neutrality is shown in Fig.~\ref{fig1}(c). It features a multi-peak structure, labeled by an increasing subscript number with decreasing peak energy as B$_1$ through B$_3$. These peaks, located about $110$ to $150$~meV below the ML peak X, correspond to phonon sidebands of momentum-indirect $KQ$ excitons \cite{Lindlau2018,Aslan2020,Forste2020}. Due to a marginal departure from the charge neutrality condition, the lowest-energy peak B$^-$ is weakly present as luminescence from bilayer $KQ$ trions \cite{Lindlau2018} with the excess electrons accommodated at the conduction band edge by the $Q$-valleys.

Interestingly, whereas the temperature-induced energy shifts are similar for the BL case, the evolution of the PL intensity with increasing temperature differs significantly from the ML. This difference is closely related to the indirect bandgap of the BL that is substantially smaller than the energy of the direct $KK$ transition. The thermal energy on the order of $10$~meV accessed in our experiments is far from sufficient to provide substantial thermal excitation of the population from the momentum-dark $KQ$ reservoir into bright $KK$ states with efficient radiative decay pathways. Therefore, the relative intensities of all peaks remain largely the same with a moderate decrease of the total emission signal. Only B$^-$ quenches strongly, as expected for a weakly bound trion state.

Following the observation and identification of the relevant transitions, we now focus on their characteristic spectral lineshapes. In the left panels of Fig.~\ref{fig2}(a) and (b), the spectra recorded at different temperatures were normalized and shifted to the energy of X and B$_{1}$ for ML and BL, respectively.
Direct comparison highlights the main difference in their spectral profiles and the associated temperature dependence. Whereas the ML bright exciton peak X is well described by a symmetric peak function such as a Lorentzian, the BL peaks are asymmetric with a pronounced high-energy shoulder (similar asymmetry is also found for the ML peaks M$_1$ through M$_4$ as discussed below). Their temperature dependence also differs from that of the bright exciton X which broadens thermally with increasing temperature and remains symmetric, while the asymmetry of the BL peaks becomes increasingly prominent. 

Clearly, the scenario of linewidth broadening due to exciton-phonon scattering applicable to X \cite{Arora2015,Selig2016} is insufficient to explain both the presence and the temperature-dependent increase of the higher-energy shoulder of BL peaks. As opposed to the direct radiative decay pathway of X, recombination of momentum-indirect excitons requires both a photon and a phonon to simultaneously satisfy energy and momentum conservation constraints. Since an arbitrary crystal momentum can be transferred to a phonon, all excitons, including those with finite center-of-mass momenta, can recombine via phonon-assisted pathways indicated in the schematics of Fig.~\ref{fig2}(c). In consequence, the thermal distribution of exciton momenta directly manifests in the characteristic high-energy shoulder of the resulting PL spectra as illustrated in Fig.~\ref{fig2}(d) and observed earlier on conventional semiconductors \cite{Umlauff1998,Xu2006,Klingshirn2007}. This is in stark contrast to direct exciton recombination (also known as zero-phonon line) where only excitons with vanishing center-of-mass momenta within the radiative cone can recombine. This difference becomes even more apparent by rescaling the energy axis at each temperature by the thermal energy of the lattice $k_{B}T$, with the high-energy shoulder shown in the right panels of Fig.~\ref{fig2}(a) and (b). In this representation, the PL spectra of X are all distinct due to symmetric thermal broadening. On the contrary, the BL spectra collapse onto a universal curve for all temperatures above $10$~K, when the thermally activated high-energy flank dominates over symmetric broadening.

To analyze these observations quantitatively, we model the phonon sidebands by Lorentzians weighted with the thermal occupation of exciton states with finite kinetic energies $E_{\mathrm{kin}}=E-E_{0}>0$, as illustrated in Fig.~\ref{fig2}(d). This approach yields the spectral lineshape $I(E)$ of phonon sideband resonances at temperature $T$ as: 
\begin{equation}
I(E)= I_0 \int_0^{\infty} \frac{\Gamma/2}{(E-E_{0})^2+(\Gamma/2)^2} \exp\left(-\frac{E_{\mathrm{kin}}}{k_B T} \right) dE, 
\label{eq1}
\end{equation}
where $I_0$ is a normalization constant, $\Gamma$ is the full-width at half-maximum (FWHM) Lorentzian linewidth in units of energy, and $k_B$ is the Boltzmann constant. The exponential term in the integral corresponds to the Boltzmann occupation of exciton states with kinetic energy measured with respect to the corresponding exciton band minimum at $E_0$ (e.\,g. as in Fig.~\ref{fig2}(c) where $E_0=E_{KQ}$). This expression for the spectral lineshape, equivalent to the result of a rigorous derivation in Ref.~\cite{Brem2020}, was used to model the experimental spectra at different temperatures using $I_0$, $E_0$, $\Gamma$ and the effective exciton temperature $T= T_{\mathrm{exc}}$ as fit parameters. We also note that the Boltzmann approximation for bosonic exciton quasiparticles should be reasonable due to rather low population densities in our experiments. 

\begin{figure}[t!]
\centering
\includegraphics[scale=1.0]{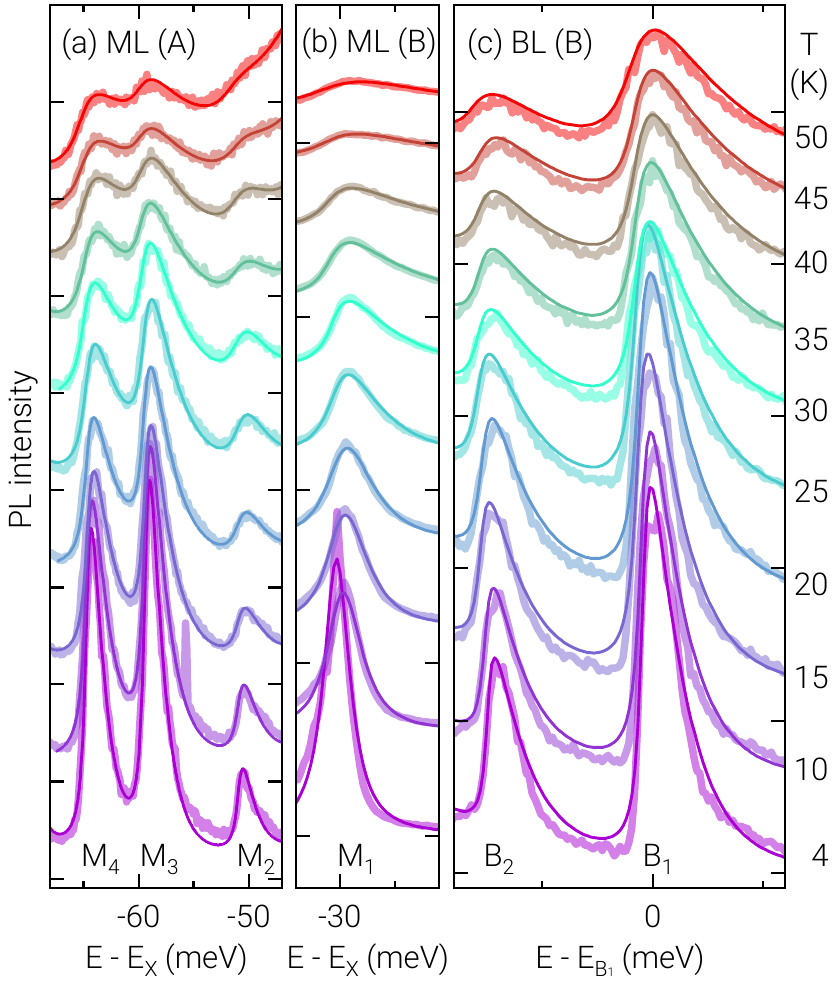}
\caption{Model fits of asymmetric peaks in (a) and (b) for ML of sample A. and B, respectively, and (c) BL of sample B. The fits (solid lines) to the spectra for temperatures ranging from $4$ to $50$~K were obtained according to Eq.~\ref{eq1} with exciton temperature, Lorentzian linewidth, peak energy and intensity as free fit parameters. Note the pronounced asymmetry on the high-energy side of all ML and BL peaks and the good agreement with the model of contributions from thermally activated excitons with finite center-of-mass momenta.}
\label{fig3}
\end{figure}

The results of least square numerical fits to the spectra of selected ML and BL peaks in samples A and B, respectively, are shown as solid lines in Fig.~\ref{fig3}(a) - (c). The ML and BL spectra were shifted at each temperature to the energy of the X and B$_1$ peaks, respectively, normalized and offset for clarity. The overall good fit quality over the studied temperature range shows that the peaks M$_{2}$, M$_{3}$ and M$_{4}$ in Fig.~\ref{fig3}(a) (note that the peaks M$_{3}$ and M$_{4}$ of sample A were used for a similar analysis in the context of exciton diffusion \cite{Wagner2021}), M$_{1}$ in Fig.~\ref{fig3}(b) as well as B$_{1}$ and B$_{2}$ in Fig.~\ref{fig3}(c) are all asymmetrically broadened and adequately modeled by Eq.~\ref{eq1}. In particular, it demonstrates the necessity to convolute a thermally broadened Lorentzian with a high-energy shoulder of thermal exciton occupation to account for the experimental observations.

\begin{figure}
\centering
\includegraphics[scale=1.0]{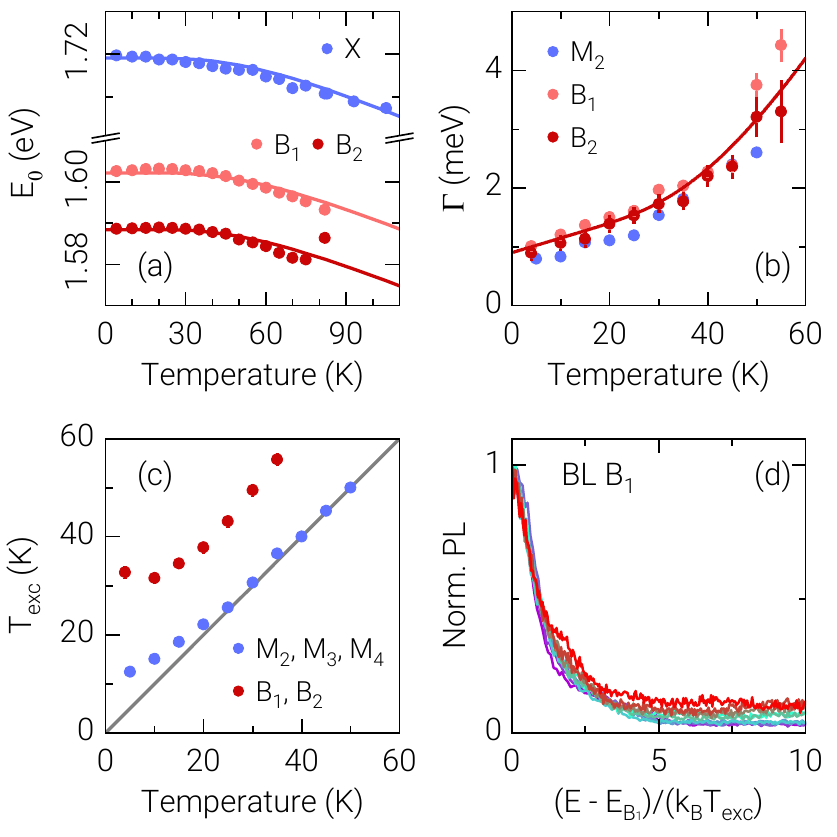}
\caption{Least-square parameters obtained from model fits. (a) Evolution of the peak energy with temperature for X in ML as well as for B$_1$ and B$_2$ in BL of sample B. The solid lines show best fits to thermal bandgap renormalization according to Eq.~\ref{eq2} with parameters from Ref.~\cite{Arora2015}. (b) Lorentzian linewidth of the ML peak M$_2$ in sample A and BL peaks B$_1$ and B$_2$ in sample B. The solid line shows best fit for BL peaks taking into account thermal phonon broadening according to Eq.~\ref{eq3}. (c) Effective exciton temperature $T_{\mathrm{exc}}$ obtained from best fits to temperature-dependent spectra with Eq.~\ref{eq1} and plotted versus the nominal sample temperature (the solid grey line shows the lattice temperature; ML and BL peaks from sample A and B, respectively). (d) Blue shoulder of B$_1$ with energy axis normalized by the effective exciton temperature $k_B T_{\mathrm{exc}}$ for direct comparison with the right panel of Fig.~\ref{fig2}(b). Note the universality of the high-energy shoulder asymmetry arising from the phonon sideband emission by thermally activated excitons.}
\label{fig4}
\end{figure}

In the progress of our analysis, we first confirm that the lattice temperature corresponds to the externally controlled temperature. To this end, we plot in Fig.~\ref{fig4}(a) the energy $E_0$ obtained from best fits to the peaks X, B$_1$ and B$_2$ in sample B as a function of temperature. The evolution shows for all peaks the expected red-shift with increasing temperature and substantial contributions from bandgap renormalization by thermally activated phonons. 
Using the relation \cite{ODonnell1991}:
\begin{equation}
E_0(T)=E_0^0-S\,\hbar\bar{\Omega}\,\left[\coth\left(\frac{\hbar\bar{\Omega}}{2k_BT}\right)-1 \right]
\label{eq2}
\end{equation}
we model the thermal energy dispersions in Fig.~\ref{fig4}(b) with the zero temperature limit of the peak energy $E_0^0$ as a free fit parameter and use the values for the coupling constant $S$ ($2.06$ and $1.75$ in ML and BL) and the average phonon energy $\hbar\bar{\Omega}=15$~meV determined previously in a much wider temperature range \cite{Arora2015}. The quantitative agreement with the observed shifts in our samples strongly supports a one-to-one correspondence between the experimental temperature and the temperature of the crystal lattice. 

The second temperature-dependent effect is the symmetric spectral broadening by pure dephasing, represented by the Lorentzian linewidth parameter $\Gamma$ in Eq.~\ref{eq1}. 
The resulting temperature dependence shown in Fig.~\ref{fig4}(b) is known to stem from thermal broadening by both linear-acoustic and fixed-energy phonons. The latter can generally include both optical and high-momentum acoustic modes in TMD materials \cite{Arora2015, Selig2016,Dey2016,Shree2018}. The approximate expression commonly used for quantitative assessment relates the spectral broadening to the phonon occupation weighted by their interaction strength with electron-hole pairs \cite{Rudin1990}:
\begin{equation}
\Gamma(T)=\Gamma_0-a\,T+b\,\left[\exp\left(\frac{\hbar\Omega}{kT}\right)-1\right]^{-1}.
\label{eq3}
\end{equation}
Here $\Gamma_{0}$ represents the zero temperature limit (excluding inhomogeneity), and $a$ and $b$ denote the coupling constants for linear-acoustic and fixed-energy phonons, respectively. The energy of the dominant phonon in the last term responsible for a super-linear linewidth dependence on the temperature is denoted by $\hbar\Omega$.

Best unconstrained fit, shown as the red solid line in Fig.~\ref{fig4}(b), was obtained with $\Gamma_0=0.9 \pm 0.2$~meV, $a=25 \pm 10~\mu$eV/K, $b=32\pm17$~meV and $\hbar\Omega=15$~meV (constraining $\hbar\Omega$ to the energy of the LO phonon $31.25$~meV instead yields a rather poor match to the data). This characteristic phonon energy is consistent with $16.3$~meV determined from coherent nonlinear spectroscopy for ML WSe$_2$ \cite{Dey2016} and close to the zone-edge phonon energy of $20$~meV in related studies of WS$_2$ \cite{Selig2016}. The coupling constant $a$ obtained from the above analysis is within the range of previously reported values of $20 - 60~\mu$eV/K for WSe$_2$ MLs \cite{Moody2015,Brem2019,Chellappan2018}. 

As the last fit parameter, we discuss the effective exciton temperature \Tx plotted against the nominal experimental temperature in Fig.~\ref{fig4}(c). From our fit procedure according to Eq.~\ref{eq1}, we find that the extracted exciton temperature generally follows that of the lattice. The agreement is very close for ML peaks in particular, as illustrated by the gray solid line in Fig.~\ref{fig4}(c) showing the lattice temperature calibrated earlier. In contrast, the momentum-indirect excitons in the BL are overheated by about $20$~K with respect to the lattice temperature.

This observation is indicative of vastly different timescales of the exciton lifetimes in the studied samples. In the ML case, the phonon sidebands M$_2$, M$_3$ and M$_4$ stem from dark excitons that decay on the order of several $100$'s of ps. As a consequence, they live sufficiently long to ensure cooling of hot excitons on a timescale of a few $10$'s of ps \cite{Rosati2020}. Similar arguments apply to the peak M$_1$ without unambiguous discrimination by the exciton $g$-factor \cite{He2020,Liu2020,Forste2020}, be it related to direct decay of momentum-indirect $KK'$ excitons \cite{He2020,Liu2020} or a phonon sideband of the $KQ$ reservoir \cite{Lindlau2017ML}. In contrast, the lifetime of the lowest-energy $KQ$ exciton reservoir in BL with phonon sidebands B$_1$ through B$_3$ is only $25$~ps \cite{Wang2014}. This difference of at least one order of magnitude in the momentum-dark exciton lifetimes in ML and BL, paired with a higher relative injection energy for the latter, accounts for overheating of momentum-dark $KQ$ excitons in the BL case. This fact is also reflected by the departure from one universal spectrum in the right panel of Fig.~\ref{fig2}(b). If, however, the blue shoulder of B$_1$ is rescaled by $k_B T_{\mathrm{exc}}$ as in Fig.~\ref{fig4}(d), we recover the expected notion that phonon sidebands reflect the thermal Boltzmann occupation of exciton states according to their kinetic energy and temperature. 

In conclusion, we experimentally identify characteristic spectral lineshape asymmetry of phonon sidebands in the PL of ML and BL WSe$_2$. In contrast to symmetric PL peaks from zero-momentum excitons, the asymmetric profiles arise from thermal distribution of excitons in long-lived dark reservoirs with finite center-of-mass momenta. The extracted exciton temperature is found to generally follow that of the crystal lattice. However, we also find deviations that are indicative of the presence of overheated exciton populations in the system. These are mostly pronounced at lowest temperatures and in the BL case, when the recombination lifetime approaches the time-scale of the cooling processes. Such direct experimental access to the temperature of excitons with finite center-of-mass momenta provides additional insight into the fundamental electron-hole recombination processes in atomically thin semiconductors. For the entire class of TMD semiconductors and heterostructures, our work emphasizes the distinctions of momentum-direct and momentum-indirect recombination pathways, identifying spectroscopic means of distinguishing the respective exciton reservoirs and highlighting the role of the phonon-assisted emission for long-lived excitons.

\section{Acknowledgements}
The authors thank Ermin Malic and members of his group for helpful discussions.
This research was funded by the European Research Council (ERC) under the Grant Agreement No.~772195 as well as the Deutsche Forschungsgemeinschaft (DFG, German Research Foundation) within the Priority Programme SPP~2244 "2DMP", Emmy Noether Initiative (CH 1672/1), SFB 1244 (project B05), and the Germany's Excellence Strategy EXC-2111-390814868. A.\,H. acknowledges support from the Center for NanoScience (CeNS) and the LMUinnovativ project Functional Nanosystems (FuNS). K.\,W. and T.\,T. acknowledge support from the Elemental Strategy Initiative conducted by the MEXT, Japan, Grant Number JPMXP0112101001, JSPS KAKENHI Grant Numbers JP20H00354 and the CREST (JPMJCR15F3), JST.


\begin{thebibliography}{56}%
\makeatletter
\providecommand \@ifxundefined [1]{%
 \@ifx{#1\undefined}
}%
\providecommand \@ifnum [1]{%
 \ifnum #1\expandafter \@firstoftwo
 \else \expandafter \@secondoftwo
 \fi
}%
\providecommand \@ifx [1]{%
 \ifx #1\expandafter \@firstoftwo
 \else \expandafter \@secondoftwo
 \fi
}%
\providecommand \natexlab [1]{#1}%
\providecommand \enquote  [1]{``#1''}%
\providecommand \bibnamefont  [1]{#1}%
\providecommand \bibfnamefont [1]{#1}%
\providecommand \citenamefont [1]{#1}%
\providecommand \href@noop [0]{\@secondoftwo}%
\providecommand \href [0]{\begingroup \@sanitize@url \@href}%
\providecommand \@href[1]{\@@startlink{#1}\@@href}%
\providecommand \@@href[1]{\endgroup#1\@@endlink}%
\providecommand \@sanitize@url [0]{\catcode `\\12\catcode `\$12\catcode
  `\&12\catcode `\#12\catcode `\^12\catcode `\_12\catcode `\%12\relax}%
\providecommand \@@startlink[1]{}%
\providecommand \@@endlink[0]{}%
\providecommand \url  [0]{\begingroup\@sanitize@url \@url }%
\providecommand \@url [1]{\endgroup\@href {#1}{\urlprefix }}%
\providecommand \urlprefix  [0]{URL }%
\providecommand \Eprint [0]{\href }%
\providecommand \doibase [0]{https://doi.org/}%
\providecommand \selectlanguage [0]{\@gobble}%
\providecommand \bibinfo  [0]{\@secondoftwo}%
\providecommand \bibfield  [0]{\@secondoftwo}%
\providecommand \translation [1]{[#1]}%
\providecommand \BibitemOpen [0]{}%
\providecommand \bibitemStop [0]{}%
\providecommand \bibitemNoStop [0]{.\EOS\space}%
\providecommand \EOS [0]{\spacefactor3000\relax}%
\providecommand \BibitemShut  [1]{\csname bibitem#1\endcsname}%
\let\auto@bib@innerbib\@empty
%
\bibitem [{\citenamefont {Macfarlane}\ \emph {et~al.}(1957)\citenamefont
  {Macfarlane}, \citenamefont {McLean}, \citenamefont {Quarrington},\ and\
  \citenamefont {Roberts}}]{MacfarlaneGe}%
  \BibitemOpen
  \bibfield  {author} {\bibinfo {author} {\bibfnamefont {G.~G.}\ \bibnamefont
  {Macfarlane}}, \bibinfo {author} {\bibfnamefont {T.~P.}\ \bibnamefont
  {McLean}}, \bibinfo {author} {\bibfnamefont {J.~E.}\ \bibnamefont
  {Quarrington}},\ and\ \bibinfo {author} {\bibfnamefont {V.}~\bibnamefont
  {Roberts}},\ }\bibfield  {title} {\bibinfo {title} {Fine structure in the
  absorption-edge spectrum of ge},\ }\href
  {https://doi.org/10.1103/PhysRev.108.1377} {\bibfield  {journal} {\bibinfo
  {journal} {Phys. Rev.}\ }\textbf {\bibinfo {volume} {108}},\ \bibinfo {pages}
  {1377} (\bibinfo {year} {1957})}\BibitemShut {NoStop}%
\bibitem [{\citenamefont {Macfarlane}\ \emph {et~al.}(1958)\citenamefont
  {Macfarlane}, \citenamefont {McLean}, \citenamefont {Quarrington},\ and\
  \citenamefont {Roberts}}]{MacfarlaneSi}%
  \BibitemOpen
  \bibfield  {author} {\bibinfo {author} {\bibfnamefont {G.~G.}\ \bibnamefont
  {Macfarlane}}, \bibinfo {author} {\bibfnamefont {T.~P.}\ \bibnamefont
  {McLean}}, \bibinfo {author} {\bibfnamefont {J.~E.}\ \bibnamefont
  {Quarrington}},\ and\ \bibinfo {author} {\bibfnamefont {V.}~\bibnamefont
  {Roberts}},\ }\bibfield  {title} {\bibinfo {title} {Fine structure in the
  absorption-edge spectrum of si},\ }\href
  {https://doi.org/10.1103/PhysRev.111.1245} {\bibfield  {journal} {\bibinfo
  {journal} {Phys. Rev.}\ }\textbf {\bibinfo {volume} {111}},\ \bibinfo {pages}
  {1245} (\bibinfo {year} {1958})}\BibitemShut {NoStop}%
\bibitem [{\citenamefont {Haynes}\ and\ \citenamefont
  {Briggs}(1952)}]{Haynes1952}%
  \BibitemOpen
  \bibfield  {author} {\bibinfo {author} {\bibfnamefont {J.}~\bibnamefont
  {Haynes}}\ and\ \bibinfo {author} {\bibfnamefont {H.}~\bibnamefont
  {Briggs}},\ }\bibfield  {title} {\bibinfo {title} {Radiation produced in
  germanium and silicon by electron-hole recombination},\ }\href@noop {}
  {\bibfield  {journal} {\bibinfo  {journal} {Phys. Rev.}\ }\textbf {\bibinfo
  {volume} {86}},\ \bibinfo {pages} {647} (\bibinfo {year} {1952})}\BibitemShut
  {NoStop}%
\bibitem [{\citenamefont {Newman}(1953)}]{Newman1953}%
  \BibitemOpen
  \bibfield  {author} {\bibinfo {author} {\bibfnamefont {R.}~\bibnamefont
  {Newman}},\ }\bibfield  {title} {\bibinfo {title} {Optical studies of
  injected carriers. ii. recombination radiation in germanium},\ }\href
  {https://doi.org/10.1103/PhysRev.91.1313} {\bibfield  {journal} {\bibinfo
  {journal} {Phys. Rev.}\ }\textbf {\bibinfo {volume} {91}},\ \bibinfo {pages}
  {1313} (\bibinfo {year} {1953})}\BibitemShut {NoStop}%
\bibitem [{\citenamefont {Haynes}(1955)}]{HaynesGe}%
  \BibitemOpen
  \bibfield  {author} {\bibinfo {author} {\bibfnamefont {J.~R.}\ \bibnamefont
  {Haynes}},\ }\bibfield  {title} {\bibinfo {title} {New radiation resulting
  from recombination of holes and electrons in germanium},\ }\href
  {https://doi.org/10.1103/PhysRev.98.1866} {\bibfield  {journal} {\bibinfo
  {journal} {Phys. Rev.}\ }\textbf {\bibinfo {volume} {98}},\ \bibinfo {pages}
  {1866} (\bibinfo {year} {1955})}\BibitemShut {NoStop}%
\bibitem [{\citenamefont {Haynes}\ and\ \citenamefont
  {Westphal}(1956)}]{HaynesSi}%
  \BibitemOpen
  \bibfield  {author} {\bibinfo {author} {\bibfnamefont {J.~R.}\ \bibnamefont
  {Haynes}}\ and\ \bibinfo {author} {\bibfnamefont {W.~C.}\ \bibnamefont
  {Westphal}},\ }\bibfield  {title} {\bibinfo {title} {Radiation resulting from
  recombination of holes and electrons in silicon},\ }\href
  {https://doi.org/10.1103/PhysRev.101.1676} {\bibfield  {journal} {\bibinfo
  {journal} {Phys. Rev.}\ }\textbf {\bibinfo {volume} {101}},\ \bibinfo {pages}
  {1676} (\bibinfo {year} {1956})}\BibitemShut {NoStop}%
\bibitem [{\citenamefont {Bonfanti}\ \emph {et~al.}(2008)\citenamefont
  {Bonfanti}, \citenamefont {Grilli}, \citenamefont {Guzzi}, \citenamefont
  {Virgilio}, \citenamefont {Grosso}, \citenamefont {Chrastina}, \citenamefont
  {Isella}, \citenamefont {von K{\"{a}}nel},\ and\ \citenamefont
  {Neels}}]{Bonfanti2008}%
  \BibitemOpen
  \bibfield  {author} {\bibinfo {author} {\bibfnamefont {M.}~\bibnamefont
  {Bonfanti}}, \bibinfo {author} {\bibfnamefont {E.}~\bibnamefont {Grilli}},
  \bibinfo {author} {\bibfnamefont {M.}~\bibnamefont {Guzzi}}, \bibinfo
  {author} {\bibfnamefont {M.}~\bibnamefont {Virgilio}}, \bibinfo {author}
  {\bibfnamefont {G.}~\bibnamefont {Grosso}}, \bibinfo {author} {\bibfnamefont
  {D.}~\bibnamefont {Chrastina}}, \bibinfo {author} {\bibfnamefont
  {G.}~\bibnamefont {Isella}}, \bibinfo {author} {\bibfnamefont
  {H.}~\bibnamefont {von K{\"{a}}nel}},\ and\ \bibinfo {author} {\bibfnamefont
  {A.}~\bibnamefont {Neels}},\ }\bibfield  {title} {\bibinfo {title} {{Optical
  transitions in Ge/SiGe multiple quantum wells with Ge-rich barriers}},\
  }\href {https://doi.org/10.1103/PhysRevB.78.041407} {\bibfield  {journal}
  {\bibinfo  {journal} {Phys. Rev. B}\ }\textbf {\bibinfo {volume} {78}},\
  \bibinfo {pages} {041407} (\bibinfo {year} {2008})}\BibitemShut {NoStop}%
\bibitem [{\citenamefont {Giorgioni}\ \emph {et~al.}(2012)\citenamefont
  {Giorgioni}, \citenamefont {Gatti}, \citenamefont {Grilli}, \citenamefont
  {Chernikov}, \citenamefont {Chatterjee}, \citenamefont {Chrastina},
  \citenamefont {Isella},\ and\ \citenamefont {Guzzi}}]{Giorgioni2012}%
  \BibitemOpen
  \bibfield  {author} {\bibinfo {author} {\bibfnamefont {A.}~\bibnamefont
  {Giorgioni}}, \bibinfo {author} {\bibfnamefont {E.}~\bibnamefont {Gatti}},
  \bibinfo {author} {\bibfnamefont {E.}~\bibnamefont {Grilli}}, \bibinfo
  {author} {\bibfnamefont {A.}~\bibnamefont {Chernikov}}, \bibinfo {author}
  {\bibfnamefont {S.}~\bibnamefont {Chatterjee}}, \bibinfo {author}
  {\bibfnamefont {D.}~\bibnamefont {Chrastina}}, \bibinfo {author}
  {\bibfnamefont {G.}~\bibnamefont {Isella}},\ and\ \bibinfo {author}
  {\bibfnamefont {M.}~\bibnamefont {Guzzi}},\ }\bibfield  {title} {\bibinfo
  {title} {{Photoluminescence decay of direct and indirect transitions in
  Ge/SiGe multiple quantum wells}},\ }\href {https://doi.org/10.1063/1.3673271}
  {\bibfield  {journal} {\bibinfo  {journal} {J. Appl. Phys.}\ }\textbf
  {\bibinfo {volume} {111}},\ \bibinfo {pages} {013501} (\bibinfo {year}
  {2012})}\BibitemShut {NoStop}%
\bibitem [{\citenamefont {Mak}\ \emph {et~al.}(2010)\citenamefont {Mak},
  \citenamefont {Lee}, \citenamefont {Hone}, \citenamefont {Shan},\ and\
  \citenamefont {Heinz}}]{Mak2010}%
  \BibitemOpen
  \bibfield  {author} {\bibinfo {author} {\bibfnamefont {K.~F.}\ \bibnamefont
  {Mak}}, \bibinfo {author} {\bibfnamefont {C.}~\bibnamefont {Lee}}, \bibinfo
  {author} {\bibfnamefont {J.}~\bibnamefont {Hone}}, \bibinfo {author}
  {\bibfnamefont {J.}~\bibnamefont {Shan}},\ and\ \bibinfo {author}
  {\bibfnamefont {T.~F.}\ \bibnamefont {Heinz}},\ }\bibfield  {title} {\bibinfo
  {title} {Atomically thin {MoS$_{2}$}: A new direct-gap semiconductor},\
  }\href {https://doi.org/10.1103/PhysRevLett.105.136805} {\bibfield  {journal}
  {\bibinfo  {journal} {Phys. Rev. Lett.}\ }\textbf {\bibinfo {volume} {105}},\
  \bibinfo {pages} {136805} (\bibinfo {year} {2010})}\BibitemShut {NoStop}%
\bibitem [{\citenamefont {Splendiani}\ \emph {et~al.}(2010)\citenamefont
  {Splendiani}, \citenamefont {Sun}, \citenamefont {Zhang}, \citenamefont {Li},
  \citenamefont {Kim}, \citenamefont {Chim}, \citenamefont {Galli},\ and\
  \citenamefont {Wang}}]{Splendiani2010}%
  \BibitemOpen
  \bibfield  {author} {\bibinfo {author} {\bibfnamefont {A.}~\bibnamefont
  {Splendiani}}, \bibinfo {author} {\bibfnamefont {L.}~\bibnamefont {Sun}},
  \bibinfo {author} {\bibfnamefont {Y.}~\bibnamefont {Zhang}}, \bibinfo
  {author} {\bibfnamefont {T.}~\bibnamefont {Li}}, \bibinfo {author}
  {\bibfnamefont {J.}~\bibnamefont {Kim}}, \bibinfo {author} {\bibfnamefont
  {C.-Y.}\ \bibnamefont {Chim}}, \bibinfo {author} {\bibfnamefont
  {G.}~\bibnamefont {Galli}},\ and\ \bibinfo {author} {\bibfnamefont
  {F.}~\bibnamefont {Wang}},\ }\bibfield  {title} {\bibinfo {title} {Emerging
  photoluminescence in monolayer {MoS$_{2}$}},\ }\href
  {https://doi.org/10.1021/nl903868w} {\bibfield  {journal} {\bibinfo
  {journal} {Nano Lett.}\ }\textbf {\bibinfo {volume} {10}},\ \bibinfo {pages}
  {1271} (\bibinfo {year} {2010})}\BibitemShut {NoStop}%
\bibitem [{\citenamefont {Wang}\ \emph {et~al.}(2018)\citenamefont {Wang},
  \citenamefont {Chernikov}, \citenamefont {Glazov}, \citenamefont {Heinz},
  \citenamefont {Marie}, \citenamefont {Amand},\ and\ \citenamefont
  {Urbaszek}}]{Wang2018}%
  \BibitemOpen
  \bibfield  {author} {\bibinfo {author} {\bibfnamefont {G.}~\bibnamefont
  {Wang}}, \bibinfo {author} {\bibfnamefont {A.}~\bibnamefont {Chernikov}},
  \bibinfo {author} {\bibfnamefont {M.~M.}\ \bibnamefont {Glazov}}, \bibinfo
  {author} {\bibfnamefont {T.~F.}\ \bibnamefont {Heinz}}, \bibinfo {author}
  {\bibfnamefont {X.}~\bibnamefont {Marie}}, \bibinfo {author} {\bibfnamefont
  {T.}~\bibnamefont {Amand}},\ and\ \bibinfo {author} {\bibfnamefont
  {B.}~\bibnamefont {Urbaszek}},\ }\bibfield  {title} {\bibinfo {title}
  {{Colloquium: Excitons in atomically thin transition metal
  dichalcogenides}},\ }\href@noop {} {\bibfield  {journal} {\bibinfo  {journal}
  {Rev. Mod. Phys.}\ }\textbf {\bibinfo {volume} {90}},\ \bibinfo {pages}
  {021001} (\bibinfo {year} {2018})}\BibitemShut {NoStop}%
\bibitem [{\citenamefont {Xu}\ \emph {et~al.}(2014)\citenamefont {Xu},
  \citenamefont {Yao}, \citenamefont {Xiao},\ and\ \citenamefont
  {Heinz}}]{Xu2014}%
  \BibitemOpen
  \bibfield  {author} {\bibinfo {author} {\bibfnamefont {X.}~\bibnamefont
  {Xu}}, \bibinfo {author} {\bibfnamefont {W.}~\bibnamefont {Yao}}, \bibinfo
  {author} {\bibfnamefont {D.}~\bibnamefont {Xiao}},\ and\ \bibinfo {author}
  {\bibfnamefont {T.~F.}\ \bibnamefont {Heinz}},\ }\bibfield  {title} {\bibinfo
  {title} {{Spin and pseudospins in layered transition metal
  dichalcogenides}},\ }\href {https://doi.org/10.1038/nphys2942} {\bibfield
  {journal} {\bibinfo  {journal} {Nat. Phys.}\ }\textbf {\bibinfo {volume}
  {10}},\ \bibinfo {pages} {343} (\bibinfo {year} {2014})}\BibitemShut
  {NoStop}%
\bibitem [{\citenamefont {Mak}\ \emph {et~al.}(2013)\citenamefont {Mak},
  \citenamefont {He}, \citenamefont {Lee}, \citenamefont {Lee}, \citenamefont
  {Hone}, \citenamefont {Heinz},\ and\ \citenamefont {Shan}}]{Mak2013}%
  \BibitemOpen
  \bibfield  {author} {\bibinfo {author} {\bibfnamefont {K.~F.}\ \bibnamefont
  {Mak}}, \bibinfo {author} {\bibfnamefont {K.}~\bibnamefont {He}}, \bibinfo
  {author} {\bibfnamefont {C.}~\bibnamefont {Lee}}, \bibinfo {author}
  {\bibfnamefont {G.~H.}\ \bibnamefont {Lee}}, \bibinfo {author} {\bibfnamefont
  {J.}~\bibnamefont {Hone}}, \bibinfo {author} {\bibfnamefont {T.~F.}\
  \bibnamefont {Heinz}},\ and\ \bibinfo {author} {\bibfnamefont
  {J.}~\bibnamefont {Shan}},\ }\bibfield  {title} {\bibinfo {title} {Tightly
  bound trions in monolayer {MoS$_{2}$}},\ }\href
  {https://doi.org/10.1038/nmat3505} {\bibfield  {journal} {\bibinfo  {journal}
  {Nat. Mater.}\ }\textbf {\bibinfo {volume} {12}},\ \bibinfo {pages} {207}
  (\bibinfo {year} {2013})}\BibitemShut {NoStop}%
\bibitem [{\citenamefont {Ross}\ \emph {et~al.}(2013)\citenamefont {Ross},
  \citenamefont {Wu}, \citenamefont {Yu}, \citenamefont {Ghimire},
  \citenamefont {Jones}, \citenamefont {Aivazian}, \citenamefont {Yan},
  \citenamefont {Mandrus}, \citenamefont {Xiao}, \citenamefont {Yao},\ and\
  \citenamefont {Xu}}]{Ross2013}%
  \BibitemOpen
  \bibfield  {author} {\bibinfo {author} {\bibfnamefont {J.~S.}\ \bibnamefont
  {Ross}}, \bibinfo {author} {\bibfnamefont {S.}~\bibnamefont {Wu}}, \bibinfo
  {author} {\bibfnamefont {H.}~\bibnamefont {Yu}}, \bibinfo {author}
  {\bibfnamefont {N.~J.}\ \bibnamefont {Ghimire}}, \bibinfo {author}
  {\bibfnamefont {A.~M.}\ \bibnamefont {Jones}}, \bibinfo {author}
  {\bibfnamefont {G.}~\bibnamefont {Aivazian}}, \bibinfo {author}
  {\bibfnamefont {J.}~\bibnamefont {Yan}}, \bibinfo {author} {\bibfnamefont
  {D.~G.}\ \bibnamefont {Mandrus}}, \bibinfo {author} {\bibfnamefont
  {D.}~\bibnamefont {Xiao}}, \bibinfo {author} {\bibfnamefont {W.}~\bibnamefont
  {Yao}},\ and\ \bibinfo {author} {\bibfnamefont {X.}~\bibnamefont {Xu}},\
  }\bibfield  {title} {\bibinfo {title} {Electrical control of neutral and
  charged excitons in a monolayer semiconductor},\ }\href@noop {} {\bibfield
  {journal} {\bibinfo  {journal} {Nat. Commun.}\ }\textbf {\bibinfo {volume}
  {4}},\ \bibinfo {pages} {1474} (\bibinfo {year} {2013})}\BibitemShut
  {NoStop}%
\bibitem [{\citenamefont {Cadiz}\ \emph {et~al.}(2017)\citenamefont {Cadiz},
  \citenamefont {Courtade}, \citenamefont {Robert}, \citenamefont {Wang},
  \citenamefont {Shen}, \citenamefont {Cai}, \citenamefont {Taniguchi},
  \citenamefont {Watanabe}, \citenamefont {Carrere}, \citenamefont {Lagarde},
  \citenamefont {Manca}, \citenamefont {Amand}, \citenamefont {Renucci},
  \citenamefont {Tongay}, \citenamefont {Marie},\ and\ \citenamefont
  {Urbaszek}}]{Cadiz2017}%
  \BibitemOpen
  \bibfield  {author} {\bibinfo {author} {\bibfnamefont {F.}~\bibnamefont
  {Cadiz}}, \bibinfo {author} {\bibfnamefont {E.}~\bibnamefont {Courtade}},
  \bibinfo {author} {\bibfnamefont {C.}~\bibnamefont {Robert}}, \bibinfo
  {author} {\bibfnamefont {G.}~\bibnamefont {Wang}}, \bibinfo {author}
  {\bibfnamefont {Y.}~\bibnamefont {Shen}}, \bibinfo {author} {\bibfnamefont
  {H.}~\bibnamefont {Cai}}, \bibinfo {author} {\bibfnamefont {T.}~\bibnamefont
  {Taniguchi}}, \bibinfo {author} {\bibfnamefont {K.}~\bibnamefont {Watanabe}},
  \bibinfo {author} {\bibfnamefont {H.}~\bibnamefont {Carrere}}, \bibinfo
  {author} {\bibfnamefont {D.}~\bibnamefont {Lagarde}}, \bibinfo {author}
  {\bibfnamefont {M.}~\bibnamefont {Manca}}, \bibinfo {author} {\bibfnamefont
  {T.}~\bibnamefont {Amand}}, \bibinfo {author} {\bibfnamefont
  {P.}~\bibnamefont {Renucci}}, \bibinfo {author} {\bibfnamefont
  {S.}~\bibnamefont {Tongay}}, \bibinfo {author} {\bibfnamefont
  {X.}~\bibnamefont {Marie}},\ and\ \bibinfo {author} {\bibfnamefont
  {B.}~\bibnamefont {Urbaszek}},\ }\bibfield  {title} {\bibinfo {title}
  {{Excitonic linewidth approaching the homogeneous limit in MoS$_2$-based van
  der waals heterostructures}},\ }\href
  {https://doi.org/10.1103/PhysRevX.7.021026} {\bibfield  {journal} {\bibinfo
  {journal} {Phys. Rev. X}\ }\textbf {\bibinfo {volume} {7}},\ \bibinfo {pages}
  {021026} (\bibinfo {year} {2017})}\BibitemShut {NoStop}%
\bibitem [{\citenamefont {Ajayi}\ \emph {et~al.}(2017)\citenamefont {Ajayi},
  \citenamefont {Ardelean}, \citenamefont {Shepard}, \citenamefont {Wang},
  \citenamefont {Antony}, \citenamefont {Taniguchi}, \citenamefont {Watanabe},
  \citenamefont {Heinz}, \citenamefont {Strauf}, \citenamefont {Zhu},\ and\
  \citenamefont {Hone}}]{Ajayi2017}%
  \BibitemOpen
  \bibfield  {author} {\bibinfo {author} {\bibfnamefont {O.~A.}\ \bibnamefont
  {Ajayi}}, \bibinfo {author} {\bibfnamefont {J.~V.}\ \bibnamefont {Ardelean}},
  \bibinfo {author} {\bibfnamefont {G.~D.}\ \bibnamefont {Shepard}}, \bibinfo
  {author} {\bibfnamefont {J.}~\bibnamefont {Wang}}, \bibinfo {author}
  {\bibfnamefont {A.}~\bibnamefont {Antony}}, \bibinfo {author} {\bibfnamefont
  {T.}~\bibnamefont {Taniguchi}}, \bibinfo {author} {\bibfnamefont
  {K.}~\bibnamefont {Watanabe}}, \bibinfo {author} {\bibfnamefont {T.~F.}\
  \bibnamefont {Heinz}}, \bibinfo {author} {\bibfnamefont {S.}~\bibnamefont
  {Strauf}}, \bibinfo {author} {\bibfnamefont {X.-Y.}\ \bibnamefont {Zhu}},\
  and\ \bibinfo {author} {\bibfnamefont {J.~C.}\ \bibnamefont {Hone}},\
  }\bibfield  {title} {\bibinfo {title} {Approaching the intrinsic
  photoluminescence linewidth in transition metal dichalcogenide monolayers},\
  }\href@noop {} {\bibfield  {journal} {\bibinfo  {journal} {2D Mater.}\
  }\textbf {\bibinfo {volume} {4}},\ \bibinfo {pages} {031011} (\bibinfo {year}
  {2017})}\BibitemShut {NoStop}%
\bibitem [{\citenamefont {Wierzbowski}\ \emph {et~al.}(2017)\citenamefont
  {Wierzbowski}, \citenamefont {Klein}, \citenamefont {Sigger}, \citenamefont
  {Straubinger}, \citenamefont {Kremser}, \citenamefont {Taniguchi},
  \citenamefont {Watanabe}, \citenamefont {Wurstbauer}, \citenamefont
  {Holleitner}, \citenamefont {Kaniber}, \citenamefont {Müller},\ and\
  \citenamefont {Finley}}]{Wierzbowski2017}%
  \BibitemOpen
  \bibfield  {author} {\bibinfo {author} {\bibfnamefont {J.}~\bibnamefont
  {Wierzbowski}}, \bibinfo {author} {\bibfnamefont {J.}~\bibnamefont {Klein}},
  \bibinfo {author} {\bibfnamefont {F.}~\bibnamefont {Sigger}}, \bibinfo
  {author} {\bibfnamefont {C.}~\bibnamefont {Straubinger}}, \bibinfo {author}
  {\bibfnamefont {M.}~\bibnamefont {Kremser}}, \bibinfo {author} {\bibfnamefont
  {T.}~\bibnamefont {Taniguchi}}, \bibinfo {author} {\bibfnamefont
  {K.}~\bibnamefont {Watanabe}}, \bibinfo {author} {\bibfnamefont
  {U.}~\bibnamefont {Wurstbauer}}, \bibinfo {author} {\bibfnamefont {A.~W.}\
  \bibnamefont {Holleitner}}, \bibinfo {author} {\bibfnamefont
  {M.}~\bibnamefont {Kaniber}}, \bibinfo {author} {\bibfnamefont
  {K.}~\bibnamefont {Müller}},\ and\ \bibinfo {author} {\bibfnamefont
  {J.~J.}\ \bibnamefont {Finley}},\ }\bibfield  {title} {\bibinfo {title}
  {Direct exciton emission from atomically thin transition metal dichalcogenide
  heterostructures near the lifetime limit},\ }\href@noop {} {\bibfield
  {journal} {\bibinfo  {journal} {Sci. Rep.}\ }\textbf {\bibinfo {volume}
  {7}},\ \bibinfo {pages} {12383} (\bibinfo {year} {2017})}\BibitemShut
  {NoStop}%
\bibitem [{\citenamefont {Zhou}\ \emph {et~al.}(2017)\citenamefont {Zhou},
  \citenamefont {Scuri}, \citenamefont {Wild}, \citenamefont {High},
  \citenamefont {Dibos}, \citenamefont {Jauregui}, \citenamefont {Shu},
  \citenamefont {{De Greve}}, \citenamefont {Pistunova}, \citenamefont {Joe},
  \citenamefont {Taniguchi}, \citenamefont {Watanabe}, \citenamefont {Kim},
  \citenamefont {Lukin},\ and\ \citenamefont {Park}}]{Zhou2017}%
  \BibitemOpen
  \bibfield  {author} {\bibinfo {author} {\bibfnamefont {Y.}~\bibnamefont
  {Zhou}}, \bibinfo {author} {\bibfnamefont {G.}~\bibnamefont {Scuri}},
  \bibinfo {author} {\bibfnamefont {D.~S.}\ \bibnamefont {Wild}}, \bibinfo
  {author} {\bibfnamefont {A.~A.}\ \bibnamefont {High}}, \bibinfo {author}
  {\bibfnamefont {A.}~\bibnamefont {Dibos}}, \bibinfo {author} {\bibfnamefont
  {L.~A.}\ \bibnamefont {Jauregui}}, \bibinfo {author} {\bibfnamefont
  {C.}~\bibnamefont {Shu}}, \bibinfo {author} {\bibfnamefont {K.}~\bibnamefont
  {{De Greve}}}, \bibinfo {author} {\bibfnamefont {K.}~\bibnamefont
  {Pistunova}}, \bibinfo {author} {\bibfnamefont {A.~Y.}\ \bibnamefont {Joe}},
  \bibinfo {author} {\bibfnamefont {T.}~\bibnamefont {Taniguchi}}, \bibinfo
  {author} {\bibfnamefont {K.}~\bibnamefont {Watanabe}}, \bibinfo {author}
  {\bibfnamefont {P.}~\bibnamefont {Kim}}, \bibinfo {author} {\bibfnamefont
  {M.~D.}\ \bibnamefont {Lukin}},\ and\ \bibinfo {author} {\bibfnamefont
  {H.}~\bibnamefont {Park}},\ }\bibfield  {title} {\bibinfo {title} {Probing
  dark excitons in atomically thin semiconductors via near-field coupling to
  surface plasmon polaritons},\ }\href@noop {} {\bibfield  {journal} {\bibinfo
  {journal} {Nat. Nanotechnol.}\ }\textbf {\bibinfo {volume} {12}},\ \bibinfo
  {pages} {856} (\bibinfo {year} {2017})}\BibitemShut {NoStop}%
\bibitem [{\citenamefont {Zhang}\ \emph {et~al.}(2017)\citenamefont {Zhang},
  \citenamefont {Cao}, \citenamefont {Lu}, \citenamefont {Lin}, \citenamefont
  {Zhang}, \citenamefont {Wang}, \citenamefont {Li}, \citenamefont {Hone},
  \citenamefont {Robinson}, \citenamefont {Smirnov}, \citenamefont {Louie},\
  and\ \citenamefont {Heinz}}]{Zhang2017}%
  \BibitemOpen
  \bibfield  {author} {\bibinfo {author} {\bibfnamefont {X.-X.}\ \bibnamefont
  {Zhang}}, \bibinfo {author} {\bibfnamefont {T.}~\bibnamefont {Cao}}, \bibinfo
  {author} {\bibfnamefont {Z.}~\bibnamefont {Lu}}, \bibinfo {author}
  {\bibfnamefont {Y.-C.}\ \bibnamefont {Lin}}, \bibinfo {author} {\bibfnamefont
  {F.}~\bibnamefont {Zhang}}, \bibinfo {author} {\bibfnamefont
  {Y.}~\bibnamefont {Wang}}, \bibinfo {author} {\bibfnamefont {Z.}~\bibnamefont
  {Li}}, \bibinfo {author} {\bibfnamefont {J.~C.}\ \bibnamefont {Hone}},
  \bibinfo {author} {\bibfnamefont {J.~A.}\ \bibnamefont {Robinson}}, \bibinfo
  {author} {\bibfnamefont {D.}~\bibnamefont {Smirnov}}, \bibinfo {author}
  {\bibfnamefont {S.~G.}\ \bibnamefont {Louie}},\ and\ \bibinfo {author}
  {\bibfnamefont {T.~F.}\ \bibnamefont {Heinz}},\ }\bibfield  {title} {\bibinfo
  {title} {Magnetic brightening and control of dark excitons in monolayer
  {WSe$_{2}$}},\ }\href {https://doi.org/10.1038/nnano.2017.10} {\bibfield
  {journal} {\bibinfo  {journal} {Nat. Nanotechnol.}\ }\textbf {\bibinfo
  {volume} {12}},\ \bibinfo {pages} {883} (\bibinfo {year} {2017})}\BibitemShut
  {NoStop}%
\bibitem [{\citenamefont {Lindlau}\ \emph {et~al.}(2017)\citenamefont
  {Lindlau}, \citenamefont {Robert}, \citenamefont {Funk}, \citenamefont
  {F{\"o}rg}, \citenamefont {Colombier}, \citenamefont {Neumann}, \citenamefont
  {Taniguchi}, \citenamefont {Watanabe}, \citenamefont {Glazov}, \citenamefont
  {Marie}, \citenamefont {Urbaszek},\ and\ \citenamefont
  {H{\"o}gele}}]{Lindlau2017ML}%
  \BibitemOpen
  \bibfield  {author} {\bibinfo {author} {\bibfnamefont {J.}~\bibnamefont
  {Lindlau}}, \bibinfo {author} {\bibfnamefont {C.}~\bibnamefont {Robert}},
  \bibinfo {author} {\bibfnamefont {V.}~\bibnamefont {Funk}}, \bibinfo {author}
  {\bibfnamefont {M.}~\bibnamefont {F{\"o}rg}}, \bibinfo {author}
  {\bibfnamefont {L.}~\bibnamefont {Colombier}}, \bibinfo {author}
  {\bibfnamefont {A.}~\bibnamefont {Neumann}}, \bibinfo {author} {\bibfnamefont
  {T.}~\bibnamefont {Taniguchi}}, \bibinfo {author} {\bibfnamefont
  {K.}~\bibnamefont {Watanabe}}, \bibinfo {author} {\bibfnamefont {M.~M.}\
  \bibnamefont {Glazov}}, \bibinfo {author} {\bibfnamefont {X.}~\bibnamefont
  {Marie}}, \bibinfo {author} {\bibfnamefont {B.}~\bibnamefont {Urbaszek}},\
  and\ \bibinfo {author} {\bibfnamefont {A.}~\bibnamefont {H{\"o}gele}},\
  }\bibfield  {title} {\bibinfo {title} {Identifying optical signatures of
  momentum-dark excitons in monolayer transition metal dichalcogenides},\
  }\href {https://arxiv.org/abs/1710.00988} {\bibfield  {journal} {\bibinfo
  {journal} {https://arxiv.org/abs/1710.00988}\ } (\bibinfo {year}
  {2017})}\BibitemShut {NoStop}%
\bibitem [{\citenamefont {Lindlau}\ \emph {et~al.}(2018)\citenamefont
  {Lindlau}, \citenamefont {Selig}, \citenamefont {Neumann}, \citenamefont
  {Colombier}, \citenamefont {F{\"o}rste}, \citenamefont {Funk}, \citenamefont
  {F{\"o}rg}, \citenamefont {Kim}, \citenamefont {Bergh{\"a}user},
  \citenamefont {Taniguchi}, \citenamefont {Watanabe}, \citenamefont {Wang},
  \citenamefont {Malic},\ and\ \citenamefont {H{\"o}gele}}]{Lindlau2018}%
  \BibitemOpen
  \bibfield  {author} {\bibinfo {author} {\bibfnamefont {J.}~\bibnamefont
  {Lindlau}}, \bibinfo {author} {\bibfnamefont {M.}~\bibnamefont {Selig}},
  \bibinfo {author} {\bibfnamefont {A.}~\bibnamefont {Neumann}}, \bibinfo
  {author} {\bibfnamefont {L.}~\bibnamefont {Colombier}}, \bibinfo {author}
  {\bibfnamefont {J.}~\bibnamefont {F{\"o}rste}}, \bibinfo {author}
  {\bibfnamefont {V.}~\bibnamefont {Funk}}, \bibinfo {author} {\bibfnamefont
  {M.}~\bibnamefont {F{\"o}rg}}, \bibinfo {author} {\bibfnamefont
  {J.}~\bibnamefont {Kim}}, \bibinfo {author} {\bibfnamefont {G.}~\bibnamefont
  {Bergh{\"a}user}}, \bibinfo {author} {\bibfnamefont {T.}~\bibnamefont
  {Taniguchi}}, \bibinfo {author} {\bibfnamefont {K.}~\bibnamefont {Watanabe}},
  \bibinfo {author} {\bibfnamefont {F.}~\bibnamefont {Wang}}, \bibinfo {author}
  {\bibfnamefont {E.}~\bibnamefont {Malic}},\ and\ \bibinfo {author}
  {\bibfnamefont {A.}~\bibnamefont {H{\"o}gele}},\ }\bibfield  {title}
  {\bibinfo {title} {The role of momentum-dark excitons in the elementary
  optical response of bilayer {WSe}$_2$},\ }\href
  {https://doi.org/10.1038/s41467-018-04877-3} {\bibfield  {journal} {\bibinfo
  {journal} {Nat. Commun.}\ }\textbf {\bibinfo {volume} {9}},\ \bibinfo {pages}
  {2586} (\bibinfo {year} {2018})}\BibitemShut {NoStop}%
\bibitem [{\citenamefont {Brem}\ \emph {et~al.}(2020)\citenamefont {Brem},
  \citenamefont {Ekman}, \citenamefont {Christiansen}, \citenamefont {Katsch},
  \citenamefont {Selig}, \citenamefont {Robert}, \citenamefont {Marie},
  \citenamefont {Urbaszek}, \citenamefont {Knorr},\ and\ \citenamefont
  {Malic}}]{Brem2020}%
  \BibitemOpen
  \bibfield  {author} {\bibinfo {author} {\bibfnamefont {S.}~\bibnamefont
  {Brem}}, \bibinfo {author} {\bibfnamefont {A.}~\bibnamefont {Ekman}},
  \bibinfo {author} {\bibfnamefont {D.}~\bibnamefont {Christiansen}}, \bibinfo
  {author} {\bibfnamefont {F.}~\bibnamefont {Katsch}}, \bibinfo {author}
  {\bibfnamefont {M.}~\bibnamefont {Selig}}, \bibinfo {author} {\bibfnamefont
  {C.}~\bibnamefont {Robert}}, \bibinfo {author} {\bibfnamefont
  {X.}~\bibnamefont {Marie}}, \bibinfo {author} {\bibfnamefont
  {B.}~\bibnamefont {Urbaszek}}, \bibinfo {author} {\bibfnamefont
  {A.}~\bibnamefont {Knorr}},\ and\ \bibinfo {author} {\bibfnamefont
  {E.}~\bibnamefont {Malic}},\ }\bibfield  {title} {\bibinfo {title}
  {{Phonon-assisted photoluminescence from indirect excitons in monolayers of
  transition-metal dichalcogenides}},\ }\href
  {https://doi.org/10.1021/acs.nanolett.0c00633} {\bibfield  {journal}
  {\bibinfo  {journal} {Nano Lett.}\ }\textbf {\bibinfo {volume} {20}},\
  \bibinfo {pages} {2849} (\bibinfo {year} {2020})}\BibitemShut {NoStop}%
\bibitem [{\citenamefont {He}\ \emph {et~al.}(2020)\citenamefont {He},
  \citenamefont {Rivera}, \citenamefont {{Van Tuan}}, \citenamefont {Wilson},
  \citenamefont {Yang}, \citenamefont {Taniguchi}, \citenamefont {Watanabe},
  \citenamefont {Yan}, \citenamefont {Mandrus}, \citenamefont {Yu},
  \citenamefont {Dery}, \citenamefont {Yao},\ and\ \citenamefont
  {Xu}}]{He2020}%
  \BibitemOpen
  \bibfield  {author} {\bibinfo {author} {\bibfnamefont {M.}~\bibnamefont
  {He}}, \bibinfo {author} {\bibfnamefont {P.}~\bibnamefont {Rivera}}, \bibinfo
  {author} {\bibfnamefont {D.}~\bibnamefont {{Van Tuan}}}, \bibinfo {author}
  {\bibfnamefont {N.~P.}\ \bibnamefont {Wilson}}, \bibinfo {author}
  {\bibfnamefont {M.}~\bibnamefont {Yang}}, \bibinfo {author} {\bibfnamefont
  {T.}~\bibnamefont {Taniguchi}}, \bibinfo {author} {\bibfnamefont
  {K.}~\bibnamefont {Watanabe}}, \bibinfo {author} {\bibfnamefont
  {J.}~\bibnamefont {Yan}}, \bibinfo {author} {\bibfnamefont {D.~G.}\
  \bibnamefont {Mandrus}}, \bibinfo {author} {\bibfnamefont {H.}~\bibnamefont
  {Yu}}, \bibinfo {author} {\bibfnamefont {H.}~\bibnamefont {Dery}}, \bibinfo
  {author} {\bibfnamefont {W.}~\bibnamefont {Yao}},\ and\ \bibinfo {author}
  {\bibfnamefont {X.}~\bibnamefont {Xu}},\ }\bibfield  {title} {\bibinfo
  {title} {{Valley phonons and exciton complexes in a monolayer
  semiconductor}},\ }\href {https://doi.org/10.1038/s41467-020-14472-0}
  {\bibfield  {journal} {\bibinfo  {journal} {Nat. Commun.}\ }\textbf {\bibinfo
  {volume} {11}},\ \bibinfo {pages} {618} (\bibinfo {year} {2020})}\BibitemShut
  {NoStop}%
\bibitem [{\citenamefont {Arora}\ \emph {et~al.}(2015)\citenamefont {Arora},
  \citenamefont {Koperski}, \citenamefont {Nogajewski}, \citenamefont {Marcus},
  \citenamefont {Faugeras},\ and\ \citenamefont {Potemski}}]{Arora2015}%
  \BibitemOpen
  \bibfield  {author} {\bibinfo {author} {\bibfnamefont {A.}~\bibnamefont
  {Arora}}, \bibinfo {author} {\bibfnamefont {M.}~\bibnamefont {Koperski}},
  \bibinfo {author} {\bibfnamefont {K.}~\bibnamefont {Nogajewski}}, \bibinfo
  {author} {\bibfnamefont {J.}~\bibnamefont {Marcus}}, \bibinfo {author}
  {\bibfnamefont {C.}~\bibnamefont {Faugeras}},\ and\ \bibinfo {author}
  {\bibfnamefont {M.}~\bibnamefont {Potemski}},\ }\bibfield  {title} {\bibinfo
  {title} {{Excitonic resonances in thin films of WSe$_2$: from monolayer to
  bulk material}},\ }\href {https://doi.org/10.1039/c5nr01536g} {\bibfield
  {journal} {\bibinfo  {journal} {Nanoscale}\ }\textbf {\bibinfo {volume}
  {7}},\ \bibinfo {pages} {10421} (\bibinfo {year} {2015})}\BibitemShut
  {NoStop}%
\bibitem [{\citenamefont {Courtade}\ \emph {et~al.}(2017)\citenamefont
  {Courtade}, \citenamefont {Semina}, \citenamefont {Manca}, \citenamefont
  {Glazov}, \citenamefont {Robert}, \citenamefont {Cadiz}, \citenamefont
  {Wang}, \citenamefont {Taniguchi}, \citenamefont {Watanabe}, \citenamefont
  {Pierre}, \citenamefont {Escoffier}, \citenamefont {Ivchenko}, \citenamefont
  {Renucci}, \citenamefont {Marie}, \citenamefont {Amand},\ and\ \citenamefont
  {Urbaszek}}]{Courtade2017}%
  \BibitemOpen
  \bibfield  {author} {\bibinfo {author} {\bibfnamefont {E.}~\bibnamefont
  {Courtade}}, \bibinfo {author} {\bibfnamefont {M.}~\bibnamefont {Semina}},
  \bibinfo {author} {\bibfnamefont {M.}~\bibnamefont {Manca}}, \bibinfo
  {author} {\bibfnamefont {M.~M.}\ \bibnamefont {Glazov}}, \bibinfo {author}
  {\bibfnamefont {C.}~\bibnamefont {Robert}}, \bibinfo {author} {\bibfnamefont
  {F.}~\bibnamefont {Cadiz}}, \bibinfo {author} {\bibfnamefont
  {G.}~\bibnamefont {Wang}}, \bibinfo {author} {\bibfnamefont {T.}~\bibnamefont
  {Taniguchi}}, \bibinfo {author} {\bibfnamefont {K.}~\bibnamefont {Watanabe}},
  \bibinfo {author} {\bibfnamefont {M.}~\bibnamefont {Pierre}}, \bibinfo
  {author} {\bibfnamefont {W.}~\bibnamefont {Escoffier}}, \bibinfo {author}
  {\bibfnamefont {E.~L.}\ \bibnamefont {Ivchenko}}, \bibinfo {author}
  {\bibfnamefont {P.}~\bibnamefont {Renucci}}, \bibinfo {author} {\bibfnamefont
  {X.}~\bibnamefont {Marie}}, \bibinfo {author} {\bibfnamefont
  {T.}~\bibnamefont {Amand}},\ and\ \bibinfo {author} {\bibfnamefont
  {B.}~\bibnamefont {Urbaszek}},\ }\bibfield  {title} {\bibinfo {title}
  {{Charged excitons in monolayer WSe$_2$: experiment and theory}},\ }\href
  {https://doi.org/10.1103/PhysRevB.96.085302} {\bibfield  {journal} {\bibinfo
  {journal} {Phys. Rev. B}\ }\textbf {\bibinfo {volume} {96}},\ \bibinfo
  {pages} {085302} (\bibinfo {year} {2017})}\BibitemShut {NoStop}%
\bibitem [{\citenamefont {Koperski}\ \emph {et~al.}(2017)\citenamefont
  {Koperski}, \citenamefont {Molas}, \citenamefont {Arora}, \citenamefont
  {Nogajewski}, \citenamefont {Slobodeniuk}, \citenamefont {Faugeras},\ and\
  \citenamefont {Potemski}}]{Koperski2017}%
  \BibitemOpen
  \bibfield  {author} {\bibinfo {author} {\bibfnamefont {M.}~\bibnamefont
  {Koperski}}, \bibinfo {author} {\bibfnamefont {M.~R.}\ \bibnamefont {Molas}},
  \bibinfo {author} {\bibfnamefont {A.}~\bibnamefont {Arora}}, \bibinfo
  {author} {\bibfnamefont {K.}~\bibnamefont {Nogajewski}}, \bibinfo {author}
  {\bibfnamefont {A.~O.}\ \bibnamefont {Slobodeniuk}}, \bibinfo {author}
  {\bibfnamefont {C.}~\bibnamefont {Faugeras}},\ and\ \bibinfo {author}
  {\bibfnamefont {M.}~\bibnamefont {Potemski}},\ }\bibfield  {title} {\bibinfo
  {title} {Optical properties of atomically thin transition metal
  dichalcogenides: Observations and puzzles},\ }\href
  {https://doi.org/10.1515/nanoph-2016-0165} {\bibfield  {journal} {\bibinfo
  {journal} {Nanophotonics}\ }\textbf {\bibinfo {volume} {6}},\ \bibinfo
  {pages} {1289} (\bibinfo {year} {2017})}\BibitemShut {NoStop}%
\bibitem [{\citenamefont {Liu}\ \emph {et~al.}(2019)\citenamefont {Liu},
  \citenamefont {van Baren}, \citenamefont {Taniguchi}, \citenamefont
  {Watanabe}, \citenamefont {Chang},\ and\ \citenamefont {Lui}}]{LiuPRR2019}%
  \BibitemOpen
  \bibfield  {author} {\bibinfo {author} {\bibfnamefont {E.}~\bibnamefont
  {Liu}}, \bibinfo {author} {\bibfnamefont {J.}~\bibnamefont {van Baren}},
  \bibinfo {author} {\bibfnamefont {T.}~\bibnamefont {Taniguchi}}, \bibinfo
  {author} {\bibfnamefont {K.}~\bibnamefont {Watanabe}}, \bibinfo {author}
  {\bibfnamefont {Y.-C.}\ \bibnamefont {Chang}},\ and\ \bibinfo {author}
  {\bibfnamefont {C.~H.}\ \bibnamefont {Lui}},\ }\bibfield  {title} {\bibinfo
  {title} {Valley-selective chiral phonon replicas of dark excitons and trions
  in monolayer {WSe}$_2$},\ }\href
  {https://doi.org/10.1103/PhysRevResearch.1.032007} {\bibfield  {journal}
  {\bibinfo  {journal} {Phys. Rev. Res.}\ }\textbf {\bibinfo {volume} {1}},\
  \bibinfo {pages} {032007} (\bibinfo {year} {2019})}\BibitemShut {NoStop}%
\bibitem [{\citenamefont {Liu}\ \emph {et~al.}(2020)\citenamefont {Liu},
  \citenamefont {van Baren}, \citenamefont {Liang}, \citenamefont {Taniguchi},
  \citenamefont {Watanabe}, \citenamefont {Gabor}, \citenamefont {Chang},\ and\
  \citenamefont {Lui}}]{Liu2020}%
  \BibitemOpen
  \bibfield  {author} {\bibinfo {author} {\bibfnamefont {E.}~\bibnamefont
  {Liu}}, \bibinfo {author} {\bibfnamefont {J.}~\bibnamefont {van Baren}},
  \bibinfo {author} {\bibfnamefont {C.-T.}\ \bibnamefont {Liang}}, \bibinfo
  {author} {\bibfnamefont {T.}~\bibnamefont {Taniguchi}}, \bibinfo {author}
  {\bibfnamefont {K.}~\bibnamefont {Watanabe}}, \bibinfo {author}
  {\bibfnamefont {N.~M.}\ \bibnamefont {Gabor}}, \bibinfo {author}
  {\bibfnamefont {Y.-C.}\ \bibnamefont {Chang}},\ and\ \bibinfo {author}
  {\bibfnamefont {C.~H.}\ \bibnamefont {Lui}},\ }\bibfield  {title} {\bibinfo
  {title} {Multipath optical recombination of intervalley dark excitons and
  trions in monolayer {WSe}$_2$},\ }\href
  {https://doi.org/10.1103/PhysRevLett.124.196802} {\bibfield  {journal}
  {\bibinfo  {journal} {Phys. Rev. Lett.}\ }\textbf {\bibinfo {volume} {124}},\
  \bibinfo {pages} {196802} (\bibinfo {year} {2020})}\BibitemShut {NoStop}%
\bibitem [{\citenamefont {F{\"{o}}rste}\ \emph {et~al.}(2020)\citenamefont
  {F{\"{o}}rste}, \citenamefont {Tepliakov}, \citenamefont {Kruchinin},
  \citenamefont {Lindlau}, \citenamefont {Funk}, \citenamefont {F{\"{o}}rg},
  \citenamefont {Watanabe}, \citenamefont {Taniguchi}, \citenamefont
  {Baimuratov},\ and\ \citenamefont {H{\"{o}}gele}}]{Forste2020}%
  \BibitemOpen
  \bibfield  {author} {\bibinfo {author} {\bibfnamefont {J.}~\bibnamefont
  {F{\"{o}}rste}}, \bibinfo {author} {\bibfnamefont {N.~V.}\ \bibnamefont
  {Tepliakov}}, \bibinfo {author} {\bibfnamefont {S.~Y.}\ \bibnamefont
  {Kruchinin}}, \bibinfo {author} {\bibfnamefont {J.}~\bibnamefont {Lindlau}},
  \bibinfo {author} {\bibfnamefont {V.}~\bibnamefont {Funk}}, \bibinfo {author}
  {\bibfnamefont {M.}~\bibnamefont {F{\"{o}}rg}}, \bibinfo {author}
  {\bibfnamefont {K.}~\bibnamefont {Watanabe}}, \bibinfo {author}
  {\bibfnamefont {T.}~\bibnamefont {Taniguchi}}, \bibinfo {author}
  {\bibfnamefont {A.~S.}\ \bibnamefont {Baimuratov}},\ and\ \bibinfo {author}
  {\bibfnamefont {A.}~\bibnamefont {H{\"{o}}gele}},\ }\bibfield  {title}
  {\bibinfo {title} {{Exciton g-factors in monolayer and bilayer WSe$_2$ from
  experiment and theory}},\ }\href {https://doi.org/10.1038/s41467-020-18019-1}
  {\bibfield  {journal} {\bibinfo  {journal} {Nat. Commun.}\ }\textbf {\bibinfo
  {volume} {11}},\ \bibinfo {pages} {4539} (\bibinfo {year}
  {2020})}\BibitemShut {NoStop}%
\bibitem [{\citenamefont {Aslan}\ \emph {et~al.}(2020)\citenamefont {Aslan},
  \citenamefont {Deng}, \citenamefont {Brongersma},\ and\ \citenamefont
  {Heinz}}]{Aslan2020}%
  \BibitemOpen
  \bibfield  {author} {\bibinfo {author} {\bibfnamefont {O.~B.}\ \bibnamefont
  {Aslan}}, \bibinfo {author} {\bibfnamefont {M.}~\bibnamefont {Deng}},
  \bibinfo {author} {\bibfnamefont {M.~L.}\ \bibnamefont {Brongersma}},\ and\
  \bibinfo {author} {\bibfnamefont {T.~F.}\ \bibnamefont {Heinz}},\ }\bibfield
  {title} {\bibinfo {title} {{Strained bilayer WSe$_2$ with reduced
  exciton-phonon coupling}},\ }\href
  {https://doi.org/10.1103/PhysRevB.101.115305} {\bibfield  {journal} {\bibinfo
   {journal} {Phys. Rev. B}\ }\textbf {\bibinfo {volume} {101}},\ \bibinfo
  {pages} {115305} (\bibinfo {year} {2020})}\BibitemShut {NoStop}%
\bibitem [{\citenamefont {Moody}\ \emph {et~al.}(2015)\citenamefont {Moody},
  \citenamefont {Kavir~Dass}, \citenamefont {Hao}, \citenamefont {Chen},
  \citenamefont {Li}, \citenamefont {Singh}, \citenamefont {Tran},
  \citenamefont {Clark}, \citenamefont {Xu}, \citenamefont {Bergh{\"a}user},
  \citenamefont {Malic}, \citenamefont {Knorr},\ and\ \citenamefont
  {Li}}]{Moody2015}%
  \BibitemOpen
  \bibfield  {author} {\bibinfo {author} {\bibfnamefont {G.}~\bibnamefont
  {Moody}}, \bibinfo {author} {\bibfnamefont {C.}~\bibnamefont {Kavir~Dass}},
  \bibinfo {author} {\bibfnamefont {K.}~\bibnamefont {Hao}}, \bibinfo {author}
  {\bibfnamefont {C.-H.}\ \bibnamefont {Chen}}, \bibinfo {author}
  {\bibfnamefont {L.-J.}\ \bibnamefont {Li}}, \bibinfo {author} {\bibfnamefont
  {A.}~\bibnamefont {Singh}}, \bibinfo {author} {\bibfnamefont
  {K.}~\bibnamefont {Tran}}, \bibinfo {author} {\bibfnamefont {G.}~\bibnamefont
  {Clark}}, \bibinfo {author} {\bibfnamefont {X.}~\bibnamefont {Xu}}, \bibinfo
  {author} {\bibfnamefont {G.}~\bibnamefont {Bergh{\"a}user}}, \bibinfo {author}
  {\bibfnamefont {E.}~\bibnamefont {Malic}}, \bibinfo {author} {\bibfnamefont
  {A.}~\bibnamefont {Knorr}},\ and\ \bibinfo {author} {\bibfnamefont
  {X.}~\bibnamefont {Li}},\ }\bibfield  {title} {\bibinfo {title} {Intrinsic
  homogeneous linewidth and broadening mechanisms of excitons in monolayer
  transition metal dichalcogenides},\ }\href
  {https://doi.org/10.1038/ncomms9315} {\bibfield  {journal} {\bibinfo
  {journal} {Nat. Commun.}\ }\textbf {\bibinfo {volume} {6}},\ \bibinfo {pages}
  {8315} (\bibinfo {year} {2015})}\BibitemShut {NoStop}%
\bibitem [{\citenamefont {Selig}\ \emph {et~al.}(2016)\citenamefont {Selig},
  \citenamefont {Bergh{\"a}user}, \citenamefont {Raja}, \citenamefont {Nagler},
  \citenamefont {Sch{\"u}ller}, \citenamefont {Heinz}, \citenamefont {Korn},
  \citenamefont {Chernikov}, \citenamefont {Malic},\ and\ \citenamefont
  {Knorr}}]{Selig2016}%
  \BibitemOpen
  \bibfield  {author} {\bibinfo {author} {\bibfnamefont {M.}~\bibnamefont
  {Selig}}, \bibinfo {author} {\bibfnamefont {G.}~\bibnamefont {Bergh{\"a}user}},
  \bibinfo {author} {\bibfnamefont {A.}~\bibnamefont {Raja}}, \bibinfo {author}
  {\bibfnamefont {P.}~\bibnamefont {Nagler}}, \bibinfo {author} {\bibfnamefont
  {C.}~\bibnamefont {Sch{\"u}ller}}, \bibinfo {author} {\bibfnamefont {T.~F.}\
  \bibnamefont {Heinz}}, \bibinfo {author} {\bibfnamefont {T.}~\bibnamefont
  {Korn}}, \bibinfo {author} {\bibfnamefont {A.}~\bibnamefont {Chernikov}},
  \bibinfo {author} {\bibfnamefont {E.}~\bibnamefont {Malic}},\ and\ \bibinfo
  {author} {\bibfnamefont {A.}~\bibnamefont {Knorr}},\ }\bibfield  {title}
  {\bibinfo {title} {Excitonic linewidth and coherence lifetime in monolayer
  transition metal dichalcogenides},\ }\href
  {https://doi.org/10.1038/ncomms13279} {\bibfield  {journal} {\bibinfo
  {journal} {Nat. Commun.}\ }\textbf {\bibinfo {volume} {7}},\ \bibinfo {pages}
  {13279} (\bibinfo {year} {2016})}\BibitemShut {NoStop}%
\bibitem [{\citenamefont {Malic}\ \emph {et~al.}(2018)\citenamefont {Malic},
  \citenamefont {Selig}, \citenamefont {Feierabend}, \citenamefont {Brem},
  \citenamefont {Christiansen}, \citenamefont {Wendler}, \citenamefont
  {Knorr},\ and\ \citenamefont {Bergh\"auser}}]{Malic2018}%
  \BibitemOpen
  \bibfield  {author} {\bibinfo {author} {\bibfnamefont {E.}~\bibnamefont
  {Malic}}, \bibinfo {author} {\bibfnamefont {M.}~\bibnamefont {Selig}},
  \bibinfo {author} {\bibfnamefont {M.}~\bibnamefont {Feierabend}}, \bibinfo
  {author} {\bibfnamefont {S.}~\bibnamefont {Brem}}, \bibinfo {author}
  {\bibfnamefont {D.}~\bibnamefont {Christiansen}}, \bibinfo {author}
  {\bibfnamefont {F.}~\bibnamefont {Wendler}}, \bibinfo {author} {\bibfnamefont
  {A.}~\bibnamefont {Knorr}},\ and\ \bibinfo {author} {\bibfnamefont
  {G.}~\bibnamefont {Bergh\"auser}},\ }\bibfield  {title} {\bibinfo {title}
  {Dark excitons in transition metal dichalcogenides},\ }\href
  {https://doi.org/10.1103/PhysRevMaterials.2.014002} {\bibfield  {journal}
  {\bibinfo  {journal} {Phys. Rev. Mater.}\ }\textbf {\bibinfo {volume} {2}},\
  \bibinfo {pages} {014002} (\bibinfo {year} {2018})}\BibitemShut {NoStop}%
\bibitem [{\citenamefont {Selig}\ \emph {et~al.}(2018)\citenamefont {Selig},
  \citenamefont {Bergh{\"a}user}, \citenamefont {Richter}, \citenamefont
  {Bratschitsch}, \citenamefont {Knorr},\ and\ \citenamefont
  {Malic}}]{Selig2018}%
  \BibitemOpen
  \bibfield  {author} {\bibinfo {author} {\bibfnamefont {M.}~\bibnamefont
  {Selig}}, \bibinfo {author} {\bibfnamefont {G.}~\bibnamefont {Bergh{\"a}user}},
  \bibinfo {author} {\bibfnamefont {M.}~\bibnamefont {Richter}}, \bibinfo
  {author} {\bibfnamefont {R.}~\bibnamefont {Bratschitsch}}, \bibinfo {author}
  {\bibfnamefont {A.}~\bibnamefont {Knorr}},\ and\ \bibinfo {author}
  {\bibfnamefont {E.}~\bibnamefont {Malic}},\ }\bibfield  {title} {\bibinfo
  {title} {Dark and bright exciton formation, thermalization, and
  photoluminescence in monolayer transition metal dichalcogenides},\ }\href
  {https://doi.org/10.1088/2053-1583/aabea3} {\bibfield  {journal} {\bibinfo
  {journal} {2D Materials}\ }\textbf {\bibinfo {volume} {5}},\ \bibinfo {pages}
  {035017} (\bibinfo {year} {2018})}\BibitemShut {NoStop}%
\bibitem [{\citenamefont {Bergh\"auser}\ \emph {et~al.}(2018)\citenamefont
  {Bergh\"auser}, \citenamefont {Steinleitner}, \citenamefont {Merkl},
  \citenamefont {Huber}, \citenamefont {Knorr},\ and\ \citenamefont
  {Malic}}]{Berghaeuser2018}%
  \BibitemOpen
  \bibfield  {author} {\bibinfo {author} {\bibfnamefont {G.}~\bibnamefont
  {Bergh\"auser}}, \bibinfo {author} {\bibfnamefont {P.}~\bibnamefont
  {Steinleitner}}, \bibinfo {author} {\bibfnamefont {P.}~\bibnamefont {Merkl}},
  \bibinfo {author} {\bibfnamefont {R.}~\bibnamefont {Huber}}, \bibinfo
  {author} {\bibfnamefont {A.}~\bibnamefont {Knorr}},\ and\ \bibinfo {author}
  {\bibfnamefont {E.}~\bibnamefont {Malic}},\ }\bibfield  {title} {\bibinfo
  {title} {Mapping of the dark exciton landscape in transition metal
  dichalcogenides},\ }\href {https://doi.org/10.1103/PhysRevB.98.020301}
  {\bibfield  {journal} {\bibinfo  {journal} {Phys. Rev. B}\ }\textbf {\bibinfo
  {volume} {98}},\ \bibinfo {pages} {020301} (\bibinfo {year}
  {2018})}\BibitemShut {NoStop}%
\bibitem [{\citenamefont {Peng}\ \emph {et~al.}(2019)\citenamefont {Peng},
  \citenamefont {Lo}, \citenamefont {Li}, \citenamefont {Huang}, \citenamefont
  {Chen}, \citenamefont {Lee}, \citenamefont {Yang},\ and\ \citenamefont
  {Cheng}}]{Peng2019}%
  \BibitemOpen
  \bibfield  {author} {\bibinfo {author} {\bibfnamefont {G.-H.}\ \bibnamefont
  {Peng}}, \bibinfo {author} {\bibfnamefont {P.-Y.}\ \bibnamefont {Lo}},
  \bibinfo {author} {\bibfnamefont {W.-H.}\ \bibnamefont {Li}}, \bibinfo
  {author} {\bibfnamefont {Y.-C.}\ \bibnamefont {Huang}}, \bibinfo {author}
  {\bibfnamefont {Y.-H.}\ \bibnamefont {Chen}}, \bibinfo {author}
  {\bibfnamefont {C.-H.}\ \bibnamefont {Lee}}, \bibinfo {author} {\bibfnamefont
  {C.-K.}\ \bibnamefont {Yang}},\ and\ \bibinfo {author} {\bibfnamefont
  {S.-J.}\ \bibnamefont {Cheng}},\ }\bibfield  {title} {\bibinfo {title}
  {{Distinctive signatures of the spin- and momentum-forbidden dark exciton
  states in the photoluminescence of strained WSe$_2$ monolayers under
  thermalization}},\ }\href {https://doi.org/10.1021/acs.nanolett.8b04786}
  {\bibfield  {journal} {\bibinfo  {journal} {Nano Lett.}\ }\textbf {\bibinfo
  {volume} {19}},\ \bibinfo {pages} {2299} (\bibinfo {year}
  {2019})}\BibitemShut {NoStop}%
\bibitem [{\citenamefont {Rosati}\ \emph {et~al.}(2020)\citenamefont {Rosati},
  \citenamefont {Wagner}, \citenamefont {Brem}, \citenamefont
  {Perea-Caus{\'{i}}n}, \citenamefont {Wietek}, \citenamefont {Zipfel},
  \citenamefont {Ziegler}, \citenamefont {Selig}, \citenamefont {Taniguchi},
  \citenamefont {Watanabe}, \citenamefont {Knorr}, \citenamefont {Chernikov},\
  and\ \citenamefont {Malic}}]{Rosati2020}%
  \BibitemOpen
  \bibfield  {author} {\bibinfo {author} {\bibfnamefont {R.}~\bibnamefont
  {Rosati}}, \bibinfo {author} {\bibfnamefont {K.}~\bibnamefont {Wagner}},
  \bibinfo {author} {\bibfnamefont {S.}~\bibnamefont {Brem}}, \bibinfo {author}
  {\bibfnamefont {R.}~\bibnamefont {Perea-Caus{\'{i}}n}}, \bibinfo {author}
  {\bibfnamefont {E.}~\bibnamefont {Wietek}}, \bibinfo {author} {\bibfnamefont
  {J.}~\bibnamefont {Zipfel}}, \bibinfo {author} {\bibfnamefont {J.~D.}\
  \bibnamefont {Ziegler}}, \bibinfo {author} {\bibfnamefont {M.}~\bibnamefont
  {Selig}}, \bibinfo {author} {\bibfnamefont {T.}~\bibnamefont {Taniguchi}},
  \bibinfo {author} {\bibfnamefont {K.}~\bibnamefont {Watanabe}}, \bibinfo
  {author} {\bibfnamefont {A.}~\bibnamefont {Knorr}}, \bibinfo {author}
  {\bibfnamefont {A.}~\bibnamefont {Chernikov}},\ and\ \bibinfo {author}
  {\bibfnamefont {E.}~\bibnamefont {Malic}},\ }\bibfield  {title} {\bibinfo
  {title} {{Temporal evolution of low-temperature phonon sidebands in
  transition metal dichalcogenides}},\ }\href
  {https://doi.org/10.1021/acsphotonics.0c00866} {\bibfield  {journal}
  {\bibinfo  {journal} {ACS Photonics}\ }\textbf {\bibinfo {volume} {7}},\
  \bibinfo {pages} {2756} (\bibinfo {year} {2020})}\BibitemShut {NoStop}%
\bibitem [{\citenamefont {Klingshirn}(2007)}]{Klingshirn2007}%
  \BibitemOpen
  \bibfield  {author} {\bibinfo {author} {\bibfnamefont {C.}~\bibnamefont
  {Klingshirn}},\ }\href@noop {} {\emph {\bibinfo {title} {{Semiconductor
  Optics}}}},\ \bibinfo {edition} {3rd}\ ed.\ (\bibinfo  {publisher} {Springer,
  Berlin Heidelberg New York},\ \bibinfo {year} {2007})\BibitemShut {NoStop}%
\bibitem [{\citenamefont {Castellanos-Gomez}\ \emph {et~al.}(2014)\citenamefont
  {Castellanos-Gomez}, \citenamefont {Buscema}, \citenamefont {Molenaar},
  \citenamefont {Singh}, \citenamefont {Janssen}, \citenamefont {{Van Der
  Zant}},\ and\ \citenamefont {Steele}}]{Castellanos-Gomez2014}%
  \BibitemOpen
  \bibfield  {author} {\bibinfo {author} {\bibfnamefont {A.}~\bibnamefont
  {Castellanos-Gomez}}, \bibinfo {author} {\bibfnamefont {M.}~\bibnamefont
  {Buscema}}, \bibinfo {author} {\bibfnamefont {R.}~\bibnamefont {Molenaar}},
  \bibinfo {author} {\bibfnamefont {V.}~\bibnamefont {Singh}}, \bibinfo
  {author} {\bibfnamefont {L.}~\bibnamefont {Janssen}}, \bibinfo {author}
  {\bibfnamefont {H.~S.}\ \bibnamefont {{Van Der Zant}}},\ and\ \bibinfo
  {author} {\bibfnamefont {G.~A.}\ \bibnamefont {Steele}},\ }\bibfield  {title}
  {\bibinfo {title} {{Deterministic transfer of two-dimensional materials by
  all-dry viscoelastic stamping}},\ }\bibfield  {journal} {\bibinfo  {journal}
  {2D Mater.}\ }\textbf {\bibinfo {volume} {1}},\ \href
  {https://doi.org/011002} {011002} (\bibinfo {year} {2014})\BibitemShut
  {NoStop}%
\bibitem [{\citenamefont {Pizzocchero}\ \emph {et~al.}(2016)\citenamefont
  {Pizzocchero}, \citenamefont {Gammelgaard}, \citenamefont {Jessen},
  \citenamefont {Caridad}, \citenamefont {Wang}, \citenamefont {Hone},
  \citenamefont {B{\o}ggild},\ and\ \citenamefont {Booth}}]{Pizzocchero2016}%
  \BibitemOpen
  \bibfield  {author} {\bibinfo {author} {\bibfnamefont {F.}~\bibnamefont
  {Pizzocchero}}, \bibinfo {author} {\bibfnamefont {L.}~\bibnamefont
  {Gammelgaard}}, \bibinfo {author} {\bibfnamefont {B.~S.}\ \bibnamefont
  {Jessen}}, \bibinfo {author} {\bibfnamefont {J.~M.}\ \bibnamefont {Caridad}},
  \bibinfo {author} {\bibfnamefont {L.}~\bibnamefont {Wang}}, \bibinfo {author}
  {\bibfnamefont {J.}~\bibnamefont {Hone}}, \bibinfo {author} {\bibfnamefont
  {P.}~\bibnamefont {B{\o}ggild}},\ and\ \bibinfo {author} {\bibfnamefont
  {T.~J.}\ \bibnamefont {Booth}},\ }\bibfield  {title} {\bibinfo {title} {{The
  hot pick-up technique for batch assembly of van der Waals
  heterostructures}},\ }\bibfield  {journal} {\bibinfo  {journal} {Nat.
  Commun.}\ }\textbf {\bibinfo {volume} {7}},\ \href {https://doi.org/11894}
  {11894} (\bibinfo {year} {2016})\BibitemShut {NoStop}%
\bibitem [{\citenamefont {Taniguchi}\ \emph {et~al.}(2018)\citenamefont
  {Taniguchi}, \citenamefont {Jain}, \citenamefont {Parzefall}, \citenamefont
  {Watanabe}, \citenamefont {Novotny}, \citenamefont {Heeg},\ and\
  \citenamefont {Bharadwaj}}]{Taniguchi2018}%
  \BibitemOpen
  \bibfield  {author} {\bibinfo {author} {\bibfnamefont {T.}~\bibnamefont
  {Taniguchi}}, \bibinfo {author} {\bibfnamefont {A.}~\bibnamefont {Jain}},
  \bibinfo {author} {\bibfnamefont {M.}~\bibnamefont {Parzefall}}, \bibinfo
  {author} {\bibfnamefont {K.}~\bibnamefont {Watanabe}}, \bibinfo {author}
  {\bibfnamefont {L.}~\bibnamefont {Novotny}}, \bibinfo {author} {\bibfnamefont
  {S.}~\bibnamefont {Heeg}},\ and\ \bibinfo {author} {\bibfnamefont
  {P.}~\bibnamefont {Bharadwaj}},\ }\bibfield  {title} {\bibinfo {title}
  {{Minimizing residues and strain in 2D materials transferred from PDMS}},\
  }\href {https://doi.org/10.1088/1361-6528/aabd90} {\bibfield  {journal}
  {\bibinfo  {journal} {Nanotechnology}\ }\textbf {\bibinfo {volume} {29}},\
  \bibinfo {pages} {265203} (\bibinfo {year} {2018})}\BibitemShut {NoStop}%
\bibitem [{\citenamefont {{Van Tuan}}\ \emph {et~al.}(2019)\citenamefont {{Van
  Tuan}}, \citenamefont {Jones}, \citenamefont {Yang}, \citenamefont {Xu},\
  and\ \citenamefont {Dery}}]{VanTuan2019}%
  \BibitemOpen
  \bibfield  {author} {\bibinfo {author} {\bibfnamefont {D.}~\bibnamefont {{Van
  Tuan}}}, \bibinfo {author} {\bibfnamefont {A.~M.}\ \bibnamefont {Jones}},
  \bibinfo {author} {\bibfnamefont {M.}~\bibnamefont {Yang}}, \bibinfo {author}
  {\bibfnamefont {X.}~\bibnamefont {Xu}},\ and\ \bibinfo {author}
  {\bibfnamefont {H.}~\bibnamefont {Dery}},\ }\bibfield  {title} {\bibinfo
  {title} {{Virtual trions in the photoluminescence of monolayer
  transition-metal dichalcogenides}},\ }\href
  {https://doi.org/10.1103/PhysRevLett.122.217401} {\bibfield  {journal}
  {\bibinfo  {journal} {Phys. Rev. Lett.}\ }\textbf {\bibinfo {volume} {122}},\
  \bibinfo {pages} {217401} (\bibinfo {year} {2019})}\BibitemShut {NoStop}%
\bibitem [{\citenamefont {Zhang}\ \emph {et~al.}(2015)\citenamefont {Zhang},
  \citenamefont {You}, \citenamefont {Zhao},\ and\ \citenamefont
  {Heinz}}]{Zhang2015}%
  \BibitemOpen
  \bibfield  {author} {\bibinfo {author} {\bibfnamefont {X.~X.}\ \bibnamefont
  {Zhang}}, \bibinfo {author} {\bibfnamefont {Y.}~\bibnamefont {You}}, \bibinfo
  {author} {\bibfnamefont {S.~Y.~F.}\ \bibnamefont {Zhao}},\ and\ \bibinfo
  {author} {\bibfnamefont {T.~F.}\ \bibnamefont {Heinz}},\ }\bibfield  {title}
  {\bibinfo {title} {{Experimental evidence for dark excitons in monolayer
  WSe$_2$}},\ }\href {https://doi.org/10.1103/PhysRevLett.115.257403}
  {\bibfield  {journal} {\bibinfo  {journal} {Phys. Rev. Lett.}\ }\textbf
  {\bibinfo {volume} {115}},\ \bibinfo {pages} {257403} (\bibinfo {year}
  {2015})}\BibitemShut {NoStop}%
\bibitem [{\citenamefont {Wang}\ \emph {et~al.}(2017)\citenamefont {Wang},
  \citenamefont {Robert}, \citenamefont {Glazov}, \citenamefont {Cadiz},
  \citenamefont {Courtade}, \citenamefont {Amand}, \citenamefont {Lagarde},
  \citenamefont {Taniguchi}, \citenamefont {Watanabe}, \citenamefont
  {Urbaszek},\ and\ \citenamefont {Marie}}]{Wang2017}%
  \BibitemOpen
  \bibfield  {author} {\bibinfo {author} {\bibfnamefont {G.}~\bibnamefont
  {Wang}}, \bibinfo {author} {\bibfnamefont {C.}~\bibnamefont {Robert}},
  \bibinfo {author} {\bibfnamefont {M.~M.}\ \bibnamefont {Glazov}}, \bibinfo
  {author} {\bibfnamefont {F.}~\bibnamefont {Cadiz}}, \bibinfo {author}
  {\bibfnamefont {E.}~\bibnamefont {Courtade}}, \bibinfo {author}
  {\bibfnamefont {T.}~\bibnamefont {Amand}}, \bibinfo {author} {\bibfnamefont
  {D.}~\bibnamefont {Lagarde}}, \bibinfo {author} {\bibfnamefont
  {T.}~\bibnamefont {Taniguchi}}, \bibinfo {author} {\bibfnamefont
  {K.}~\bibnamefont {Watanabe}}, \bibinfo {author} {\bibfnamefont
  {B.}~\bibnamefont {Urbaszek}},\ and\ \bibinfo {author} {\bibfnamefont
  {X.}~\bibnamefont {Marie}},\ }\bibfield  {title} {\bibinfo {title} {{In-plane
  propagation of light in transition metal dichalcogenide monolayers: optical
  selection rules}},\ }\href@noop {} {\bibfield  {journal} {\bibinfo  {journal}
  {Phys. Rev. Lett.}\ }\textbf {\bibinfo {volume} {119}},\ \bibinfo {pages}
  {047401} (\bibinfo {year} {2017})}\BibitemShut {NoStop}%
\bibitem [{\citenamefont {Robert}\ \emph {et~al.}(2017)\citenamefont {Robert},
  \citenamefont {Amand}, \citenamefont {Cadiz}, \citenamefont {Lagarde},
  \citenamefont {Courtade}, \citenamefont {Manca}, \citenamefont {Taniguchi},
  \citenamefont {Watanabe}, \citenamefont {Urbaszek},\ and\ \citenamefont
  {Marie}}]{Robert2017}%
  \BibitemOpen
  \bibfield  {author} {\bibinfo {author} {\bibfnamefont {C.}~\bibnamefont
  {Robert}}, \bibinfo {author} {\bibfnamefont {T.}~\bibnamefont {Amand}},
  \bibinfo {author} {\bibfnamefont {F.}~\bibnamefont {Cadiz}}, \bibinfo
  {author} {\bibfnamefont {D.}~\bibnamefont {Lagarde}}, \bibinfo {author}
  {\bibfnamefont {E.}~\bibnamefont {Courtade}}, \bibinfo {author}
  {\bibfnamefont {M.}~\bibnamefont {Manca}}, \bibinfo {author} {\bibfnamefont
  {T.}~\bibnamefont {Taniguchi}}, \bibinfo {author} {\bibfnamefont
  {K.}~\bibnamefont {Watanabe}}, \bibinfo {author} {\bibfnamefont
  {B.}~\bibnamefont {Urbaszek}},\ and\ \bibinfo {author} {\bibfnamefont
  {X.}~\bibnamefont {Marie}},\ }\bibfield  {title} {\bibinfo {title} {{Fine
  structure and lifetime of dark excitons in transition metal dichalcogenide
  monolayers}},\ }\href {https://doi.org/10.1103/PhysRevB.96.155423} {\bibfield
   {journal} {\bibinfo  {journal} {Phys. Rev. B}\ }\textbf {\bibinfo {volume}
  {96}},\ \bibinfo {pages} {155423} (\bibinfo {year} {2017})}\BibitemShut
  {NoStop}%
\bibitem [{\citenamefont {Li}\ \emph {et~al.}(2019)\citenamefont {Li},
  \citenamefont {Wang}, \citenamefont {Jin}, \citenamefont {Lu}, \citenamefont
  {Lian}, \citenamefont {Meng}, \citenamefont {Blei}, \citenamefont {Gao},
  \citenamefont {Taniguchi}, \citenamefont {Watanabe}, \citenamefont {Ren},
  \citenamefont {Tongay}, \citenamefont {Yang}, \citenamefont {Smirnov},
  \citenamefont {Cao},\ and\ \citenamefont {Shi}}]{Li2019}%
  \BibitemOpen
  \bibfield  {author} {\bibinfo {author} {\bibfnamefont {Z.}~\bibnamefont
  {Li}}, \bibinfo {author} {\bibfnamefont {T.}~\bibnamefont {Wang}}, \bibinfo
  {author} {\bibfnamefont {C.}~\bibnamefont {Jin}}, \bibinfo {author}
  {\bibfnamefont {Z.}~\bibnamefont {Lu}}, \bibinfo {author} {\bibfnamefont
  {Z.}~\bibnamefont {Lian}}, \bibinfo {author} {\bibfnamefont {Y.}~\bibnamefont
  {Meng}}, \bibinfo {author} {\bibfnamefont {M.}~\bibnamefont {Blei}}, \bibinfo
  {author} {\bibfnamefont {S.}~\bibnamefont {Gao}}, \bibinfo {author}
  {\bibfnamefont {T.}~\bibnamefont {Taniguchi}}, \bibinfo {author}
  {\bibfnamefont {K.}~\bibnamefont {Watanabe}}, \bibinfo {author}
  {\bibfnamefont {T.}~\bibnamefont {Ren}}, \bibinfo {author} {\bibfnamefont
  {S.}~\bibnamefont {Tongay}}, \bibinfo {author} {\bibfnamefont
  {L.}~\bibnamefont {Yang}}, \bibinfo {author} {\bibfnamefont {D.}~\bibnamefont
  {Smirnov}}, \bibinfo {author} {\bibfnamefont {T.}~\bibnamefont {Cao}},\ and\
  \bibinfo {author} {\bibfnamefont {S.~F.}\ \bibnamefont {Shi}},\ }\bibfield
  {title} {\bibinfo {title} {{Emerging photoluminescence from the dark-exciton
  phonon replica in monolayer WSe$_2$}},\ }\href
  {https://doi.org/10.1038/s41467-019-10477-6} {\bibfield  {journal} {\bibinfo
  {journal} {Nat. Commun.}\ }\textbf {\bibinfo {volume} {10}},\ \bibinfo
  {pages} {2469} (\bibinfo {year} {2019})}\BibitemShut {NoStop}%
\bibitem [{\citenamefont {Umlauff}\ \emph {et~al.}(1998)\citenamefont
  {Umlauff}, \citenamefont {Hoffmann}, \citenamefont {Kalt}, \citenamefont
  {Langbein}, \citenamefont {Hvam}, \citenamefont {Scholl}, \citenamefont
  {S{\"{o}}llner}, \citenamefont {Heuken}, \citenamefont {Jobst},\ and\
  \citenamefont {Hommel}}]{Umlauff1998}%
  \BibitemOpen
  \bibfield  {author} {\bibinfo {author} {\bibfnamefont {M.}~\bibnamefont
  {Umlauff}}, \bibinfo {author} {\bibfnamefont {J.}~\bibnamefont {Hoffmann}},
  \bibinfo {author} {\bibfnamefont {H.}~\bibnamefont {Kalt}}, \bibinfo {author}
  {\bibfnamefont {W.}~\bibnamefont {Langbein}}, \bibinfo {author}
  {\bibfnamefont {J.~M.}\ \bibnamefont {Hvam}}, \bibinfo {author}
  {\bibfnamefont {M.}~\bibnamefont {Scholl}}, \bibinfo {author} {\bibfnamefont
  {J.}~\bibnamefont {S{\"{o}}llner}}, \bibinfo {author} {\bibfnamefont
  {M.}~\bibnamefont {Heuken}}, \bibinfo {author} {\bibfnamefont
  {B.}~\bibnamefont {Jobst}},\ and\ \bibinfo {author} {\bibfnamefont
  {D.}~\bibnamefont {Hommel}},\ }\bibfield  {title} {\bibinfo {title} {{Direct
  observation of free-exciton thermalization in quantum-well structures}},\
  }\href {https://doi.org/10.1103/PhysRevB.57.1390} {\bibfield  {journal}
  {\bibinfo  {journal} {Phys. Rev. B}\ }\textbf {\bibinfo {volume} {57}},\
  \bibinfo {pages} {1390} (\bibinfo {year} {1998})}\BibitemShut {NoStop}%
\bibitem [{\citenamefont {Xu}\ \emph {et~al.}(2006)\citenamefont {Xu},
  \citenamefont {Li}, \citenamefont {Xiong},\ and\ \citenamefont
  {Che}}]{Xu2006}%
  \BibitemOpen
  \bibfield  {author} {\bibinfo {author} {\bibfnamefont {S.~J.}\ \bibnamefont
  {Xu}}, \bibinfo {author} {\bibfnamefont {G.~Q.}\ \bibnamefont {Li}}, \bibinfo
  {author} {\bibfnamefont {S.-J.}\ \bibnamefont {Xiong}},\ and\ \bibinfo
  {author} {\bibfnamefont {C.~M.}\ \bibnamefont {Che}},\ }\bibfield  {title}
  {\bibinfo {title} {{Temperature dependence of the LO phonon sidebands in free
  exciton emission of GaN}},\ }\href {https://doi.org/10.1063/1.2188034}
  {\bibfield  {journal} {\bibinfo  {journal} {J. Appl. Phys.}\ }\textbf
  {\bibinfo {volume} {99}},\ \bibinfo {pages} {073508} (\bibinfo {year}
  {2006})}\BibitemShut {NoStop}%
\bibitem [{\citenamefont {Wagner}\ \emph {et~al.}(2021)\citenamefont {Wagner},
  \citenamefont {Zipfel}, \citenamefont {Rosati}, \citenamefont {Wietek},
  \citenamefont {Ziegler}, \citenamefont {Brem}, \citenamefont
  {Perea-Caus\'{\i}n}, \citenamefont {Taniguchi}, \citenamefont {Watanabe},
  \citenamefont {Glazov}, \citenamefont {Malic},\ and\ \citenamefont
  {Chernikov}}]{Wagner2021}%
  \BibitemOpen
  \bibfield  {author} {\bibinfo {author} {\bibfnamefont {K.}~\bibnamefont
  {Wagner}}, \bibinfo {author} {\bibfnamefont {J.}~\bibnamefont {Zipfel}},
  \bibinfo {author} {\bibfnamefont {R.}~\bibnamefont {Rosati}}, \bibinfo
  {author} {\bibfnamefont {E.}~\bibnamefont {Wietek}}, \bibinfo {author}
  {\bibfnamefont {J.~D.}\ \bibnamefont {Ziegler}}, \bibinfo {author}
  {\bibfnamefont {S.}~\bibnamefont {Brem}}, \bibinfo {author} {\bibfnamefont
  {R.}~\bibnamefont {Perea-Caus\'{\i}n}}, \bibinfo {author} {\bibfnamefont
  {T.}~\bibnamefont {Taniguchi}}, \bibinfo {author} {\bibfnamefont
  {K.}~\bibnamefont {Watanabe}}, \bibinfo {author} {\bibfnamefont {M.~M.}\
  \bibnamefont {Glazov}}, \bibinfo {author} {\bibfnamefont {E.}~\bibnamefont
  {Malic}},\ and\ \bibinfo {author} {\bibfnamefont {A.}~\bibnamefont
  {Chernikov}},\ }\bibfield  {title} {\bibinfo {title} {Nonclassical exciton
  diffusion in monolayer {WSe$_{2}$}},\ }\href
  {https://doi.org/10.1103/PhysRevLett.127.076801} {\bibfield  {journal}
  {\bibinfo  {journal} {Phys. Rev. Lett.}\ }\textbf {\bibinfo {volume} {127}},\
  \bibinfo {pages} {076801} (\bibinfo {year} {2021})}\BibitemShut {NoStop}%
\bibitem [{\citenamefont {O'Donnell}\ and\ \citenamefont
  {Chen}(1991)}]{ODonnell1991}%
  \BibitemOpen
  \bibfield  {author} {\bibinfo {author} {\bibfnamefont {K.~P.}\ \bibnamefont
  {O'Donnell}}\ and\ \bibinfo {author} {\bibfnamefont {X.}~\bibnamefont
  {Chen}},\ }\bibfield  {title} {\bibinfo {title} {{Temperature dependence of
  semiconductor band gaps}},\ }\href {https://doi.org/10.1063/1.104723}
  {\bibfield  {journal} {\bibinfo  {journal} {Appl. Phys. Lett.}\ }\textbf
  {\bibinfo {volume} {58}},\ \bibinfo {pages} {2924} (\bibinfo {year}
  {1991})}\BibitemShut {NoStop}%
\bibitem [{\citenamefont {Dey}\ \emph {et~al.}(2016)\citenamefont {Dey},
  \citenamefont {Paul}, \citenamefont {Wang}, \citenamefont {Stevens},
  \citenamefont {Liu}, \citenamefont {Romero}, \citenamefont {Shan},
  \citenamefont {Hilton},\ and\ \citenamefont {Karaiskaj}}]{Dey2016}%
  \BibitemOpen
  \bibfield  {author} {\bibinfo {author} {\bibfnamefont {P.}~\bibnamefont
  {Dey}}, \bibinfo {author} {\bibfnamefont {J.}~\bibnamefont {Paul}}, \bibinfo
  {author} {\bibfnamefont {Z.}~\bibnamefont {Wang}}, \bibinfo {author}
  {\bibfnamefont {C.~E.}\ \bibnamefont {Stevens}}, \bibinfo {author}
  {\bibfnamefont {C.}~\bibnamefont {Liu}}, \bibinfo {author} {\bibfnamefont
  {A.~H.}\ \bibnamefont {Romero}}, \bibinfo {author} {\bibfnamefont
  {J.}~\bibnamefont {Shan}}, \bibinfo {author} {\bibfnamefont {D.~J.}\
  \bibnamefont {Hilton}},\ and\ \bibinfo {author} {\bibfnamefont
  {D.}~\bibnamefont {Karaiskaj}},\ }\bibfield  {title} {\bibinfo {title}
  {{Optical coherence in atomic-monolayer transition-metal dichalcogenides
  limited by electron-phonon interactions}},\ }\href@noop {} {\bibfield
  {journal} {\bibinfo  {journal} {Phys. Rev. Lett.}\ }\textbf {\bibinfo
  {volume} {116}},\ \bibinfo {pages} {127402} (\bibinfo {year}
  {2016})}\BibitemShut {NoStop}%
\bibitem [{\citenamefont {Shree}\ \emph {et~al.}(2018)\citenamefont {Shree},
  \citenamefont {Semina}, \citenamefont {Robert}, \citenamefont {Han},
  \citenamefont {Amand}, \citenamefont {Balocchi}, \citenamefont {Manca},
  \citenamefont {Courtade}, \citenamefont {Marie}, \citenamefont {Taniguchi},
  \citenamefont {Watanabe}, \citenamefont {Glazov},\ and\ \citenamefont
  {Urbaszek}}]{Shree2018}%
  \BibitemOpen
  \bibfield  {author} {\bibinfo {author} {\bibfnamefont {S.}~\bibnamefont
  {Shree}}, \bibinfo {author} {\bibfnamefont {M.}~\bibnamefont {Semina}},
  \bibinfo {author} {\bibfnamefont {C.}~\bibnamefont {Robert}}, \bibinfo
  {author} {\bibfnamefont {B.}~\bibnamefont {Han}}, \bibinfo {author}
  {\bibfnamefont {T.}~\bibnamefont {Amand}}, \bibinfo {author} {\bibfnamefont
  {A.}~\bibnamefont {Balocchi}}, \bibinfo {author} {\bibfnamefont
  {M.}~\bibnamefont {Manca}}, \bibinfo {author} {\bibfnamefont
  {E.}~\bibnamefont {Courtade}}, \bibinfo {author} {\bibfnamefont
  {X.}~\bibnamefont {Marie}}, \bibinfo {author} {\bibfnamefont
  {T.}~\bibnamefont {Taniguchi}}, \bibinfo {author} {\bibfnamefont
  {K.}~\bibnamefont {Watanabe}}, \bibinfo {author} {\bibfnamefont {M.~M.}\
  \bibnamefont {Glazov}},\ and\ \bibinfo {author} {\bibfnamefont
  {B.}~\bibnamefont {Urbaszek}},\ }\bibfield  {title} {\bibinfo {title}
  {{Observation of exciton-phonon coupling in MoSe$_2$ monolayers}},\ }\href
  {https://doi.org/10.1103/PhysRevB.98.035302} {\bibfield  {journal} {\bibinfo
  {journal} {Phys. Rev. B}\ }\textbf {\bibinfo {volume} {98}},\ \bibinfo
  {pages} {035302} (\bibinfo {year} {2018})}\BibitemShut {NoStop}%
\bibitem [{\citenamefont {Rudin}\ \emph {et~al.}(1990)\citenamefont {Rudin},
  \citenamefont {Reinecke},\ and\ \citenamefont {Segall}}]{Rudin1990}%
  \BibitemOpen
  \bibfield  {author} {\bibinfo {author} {\bibfnamefont {S.}~\bibnamefont
  {Rudin}}, \bibinfo {author} {\bibfnamefont {T.}~\bibnamefont {Reinecke}},\
  and\ \bibinfo {author} {\bibfnamefont {B.}~\bibnamefont {Segall}},\
  }\bibfield  {title} {\bibinfo {title} {{Temperature-dependent exciton
  linewidths in semiconductors}},\ }\href
  {https://doi.org/10.1103/PhysRevB.42.11218} {\bibfield  {journal} {\bibinfo
  {journal} {Phys. Rev. B}\ }\textbf {\bibinfo {volume} {42}},\ \bibinfo
  {pages} {11218} (\bibinfo {year} {1990})}\BibitemShut {NoStop}%
\bibitem [{\citenamefont {Brem}\ \emph {et~al.}(2019)\citenamefont {Brem},
  \citenamefont {Zipfel}, \citenamefont {Selig}, \citenamefont {Raja},
  \citenamefont {Waldecker}, \citenamefont {Ziegler}, \citenamefont
  {Taniguchi}, \citenamefont {Watanabe}, \citenamefont {Chernikov},\ and\
  \citenamefont {Malic}}]{Brem2019}%
  \BibitemOpen
  \bibfield  {author} {\bibinfo {author} {\bibfnamefont {S.}~\bibnamefont
  {Brem}}, \bibinfo {author} {\bibfnamefont {J.}~\bibnamefont {Zipfel}},
  \bibinfo {author} {\bibfnamefont {M.}~\bibnamefont {Selig}}, \bibinfo
  {author} {\bibfnamefont {A.}~\bibnamefont {Raja}}, \bibinfo {author}
  {\bibfnamefont {L.}~\bibnamefont {Waldecker}}, \bibinfo {author}
  {\bibfnamefont {J.~D.}\ \bibnamefont {Ziegler}}, \bibinfo {author}
  {\bibfnamefont {T.}~\bibnamefont {Taniguchi}}, \bibinfo {author}
  {\bibfnamefont {K.}~\bibnamefont {Watanabe}}, \bibinfo {author}
  {\bibfnamefont {A.}~\bibnamefont {Chernikov}},\ and\ \bibinfo {author}
  {\bibfnamefont {E.}~\bibnamefont {Malic}},\ }\bibfield  {title} {\bibinfo
  {title} {{Intrinsic lifetime of higher excitonic states in tungsten
  diselenide monolayers}},\ }\href {https://doi.org/10.1039/C9NR04211C}
  {\bibfield  {journal} {\bibinfo  {journal} {Nanoscale}\ }\textbf {\bibinfo
  {volume} {11}},\ \bibinfo {pages} {12381} (\bibinfo {year}
  {2019})}\BibitemShut {NoStop}%
\bibitem [{\citenamefont {Chellappan}\ \emph {et~al.}(2018)\citenamefont
  {Chellappan}, \citenamefont {Pang}, \citenamefont {Sarkar}, \citenamefont
  {Ooi},\ and\ \citenamefont {Goh}}]{Chellappan2018}%
  \BibitemOpen
  \bibfield  {author} {\bibinfo {author} {\bibfnamefont {V.}~\bibnamefont
  {Chellappan}}, \bibinfo {author} {\bibfnamefont {A.~L.~C.}\ \bibnamefont
  {Pang}}, \bibinfo {author} {\bibfnamefont {S.}~\bibnamefont {Sarkar}},
  \bibinfo {author} {\bibfnamefont {Z.~E.}\ \bibnamefont {Ooi}},\ and\ \bibinfo
  {author} {\bibfnamefont {K.~E.~J.}\ \bibnamefont {Goh}},\ }\bibfield  {title}
  {\bibinfo {title} {{Effect of phonons on valley depolarization in monolayer
  WSe$_2$}},\ }\href {https://doi.org/10.1007/s13391-018-0086-2} {\bibfield
  {journal} {\bibinfo  {journal} {Electron. Mater. Lett.}\ }\textbf {\bibinfo
  {volume} {14}},\ \bibinfo {pages} {766} (\bibinfo {year} {2018})}\BibitemShut
  {NoStop}%
\bibitem [{\citenamefont {Wang}\ \emph {et~al.}(2014)\citenamefont {Wang},
  \citenamefont {Marie}, \citenamefont {Bouet}, \citenamefont {Vidal},
  \citenamefont {Balocchi}, \citenamefont {Amand}, \citenamefont {Lagarde},\
  and\ \citenamefont {Urbaszek}}]{Wang2014}%
  \BibitemOpen
  \bibfield  {author} {\bibinfo {author} {\bibfnamefont {G.}~\bibnamefont
  {Wang}}, \bibinfo {author} {\bibfnamefont {X.}~\bibnamefont {Marie}},
  \bibinfo {author} {\bibfnamefont {L.}~\bibnamefont {Bouet}}, \bibinfo
  {author} {\bibfnamefont {M.}~\bibnamefont {Vidal}}, \bibinfo {author}
  {\bibfnamefont {A.}~\bibnamefont {Balocchi}}, \bibinfo {author}
  {\bibfnamefont {T.}~\bibnamefont {Amand}}, \bibinfo {author} {\bibfnamefont
  {D.}~\bibnamefont {Lagarde}},\ and\ \bibinfo {author} {\bibfnamefont
  {B.}~\bibnamefont {Urbaszek}},\ }\bibfield  {title} {\bibinfo {title}
  {{Exciton dynamics in WSe$_2$ bilayers}},\ }\href
  {https://doi.org/10.1063/1.4900945} {\bibfield  {journal} {\bibinfo
  {journal} {Appl. Phys. Lett.}\ }\textbf {\bibinfo {volume} {105}},\ \bibinfo
  {pages} {182105} (\bibinfo {year} {2014})}\BibitemShut {NoStop}%
\end{thebibliography}

%

\end{document}